# Supply Network Formation and Fragility†

*By* Matthew Elliott, Benjamin Golub, and Matthew V. Leduc*

*We model the production of complex goods in a large supply network. Each firm sources several essential inputs through relationships with other firms. Individual supply relationships are at risk of idiosyncratic failure, which threatens to disrupt production. To protect against this, firms multisource inputs and strategically invest to make relationships stronger, trading off the cost of investment against the benefits of increased robustness. A supply network is called fragile if aggregate output is very sensitive to small aggregate shocks. We show that supply networks of intermediate productivity are fragile in equilibrium, even though this is always inefficient. The endogenous configuration of supply networks provides a new channel for the powerful amplification of shocks.* (*JEL* D21, G31, L14)

Complex supply networks are a central feature of the modern economy. Consider a product such as an airplane. It consists of multiple parts essential for its production, many of which are customized. For instance, Rolls-Royce designed and developed its Trent 900 engine for the Airbus A380; Airbus could not just buy the engine it requires off-the-shelf. Such inputs are tailored to meet the customer's specifications, and there are often only a few potential suppliers from whom a firm can source a given type of input. Thus, a particular airplane producer is reliant on successful delivery of its inputs through particular supply relationships it has formed, rather than on the aggregate supply of a generic type of input. Many of the suppliers, in turn, are in a similar position: engaged in complex production using multiple customized inputs, and so on.[1] Due to the resulting interdependencies, idiosyncratic disruptions to production processes somewhere in a network can have far-reaching effects, causing damage that cascades through the supply chain and affects many downstream firms. Examples of such idiosyncratic shocks include a delay in shipment, a fire at a factory, a misunderstanding by a supplier resulting in the delivery of an unsuitable component, or a strike by workers.[2]

*Elliott: University of Cambridge (email: mle30@cam.ac.uk); Golub: Northwestern University (email: ben.golub@gmail.com); Leduc: Paris School of Economics and Université Paris 1 (Panthéon-Sorbonne) (email: mattvleduc@gmail.com). Sylvain Chassang was the coeditor for this article. This project has received funding from the European Research Council (ERC) under the European Union's Horizon 2020 research and innovation program (grant agreement 757229) and the JM Keynes Fellowships Fund (Elliott); the Joint Center for History and Economics; the Pershing Square Fund for Research on the Foundations of Human Behavior; and the National Science Foundation under grant SES-1629446 (Golub). We thank Joey Feffer, Riako Granzier-Nakajima, Yixi Jiang, Kostis Leledakis, Thomas Pellet, Rithvik Rao, and Brit Sharoni for excellent research assistance. We thank numerous seminar participants for their comments. For helpful conversations we are grateful to Daron Acemoglu, Nageeb Ali, Pol Antrás, David Baqaee, Vasco Carvalho, Olivier Compte, Krishna Dasaratha, Selman Erol, Marcel Fafchamps, Emmanuel Farhi, Alex Frankel, Sanjeev Goyal, Philippe Jehiel, Matthew O. Jackson, Chad Jones, Annie Liang, Eric Maskin, Marc Melitz, Marzena Rostek, Emma Rothschild, Jesse Shapiro, Ludwig Straub, Alireza Tahbaz-Salehi, Larry Samuelson, Andrzej Skrzypacz, Juuso Välimäki and Rakesh Vohra. We are grateful to two anonymous referees and the coeditor for their constructive guidance in revising the article for publication.
We dedicate this article to the memory of Emmanuel Farhi.
†Go to https://doi.org/10.1257/aer.20210220 to visit the article page for additional materials and author disclosure statements.
[1] An Airbus A380 has millions of parts produced by more than a thousand companies (Slutsken 2018).

We develop a theory in which firms insure against supply disruptions by strategically investing in relationships with potential suppliers, trading off private gains in the robustness of their production against the cost of maintaining strong supply relationships. Our main results examine how the resulting equilibrium supply networks respond to idiosyncratic and aggregate risk. We find that, in equilibrium, (i) the economy is robust to idiosyncratic shocks, yet (ii) small shocks that systemically affect the functioning of supply relationships are massively amplified. Moreover, (iii) the functioning of many unrelated supply chains is highly correlated, and (iv) the complexity of production is key to the nature of these effects and the level of aggregate volatility.

Underlying these results is a discontinuous phase transition in the structure of production networks that arises due to production being complex—reliant on multiple inputs at many stages. Thus, a theoretical contribution of our work is the study of novel equilibrium fragilities in the strategic formation of large supply networks, along with new methods for analyzing them.

We now describe a simple model of interfirm sourcing relationships and their disruption by shocks. There are many products (e.g., airplanes, engines, etc.). Each has many differentiated varieties, produced by small, specialized firms. A given product has a set of customized inputs that must be sourced via supply relationships and that are essential to its production—e.g., an airplane requires engines, navigation systems, etc. To source these specific, compatible varieties, firms may *multisource*, maintaining several substitutable sourcing options. Each of a firm's potential supply relationships may operate successfully or not: e.g., one engine manufacturer's delivery may be delayed by a strike, while another is able to deliver normally. For a firm to be functional, it must have, for each of its essential inputs, at least one operating supply relationship to a firm producing that input. These producers must in turn satisfy the same condition to be functional, and so on—until a point in the supply chain where no customized inputs are required. Our modeling of supply relationship risk is simple: independently, each relationship avoids a bad shock with a probability called the *relationship strength*.[3] This probability represents the chance of avoiding logistical disruptions and failures of contracts in any specific relationship. Given the set of operational relationships, the firms that are able to produce then purchase their required inputs and sell their products to other firms as well as to consumers. Social welfare is increasing in the number of firms able to produce. Firms earn profits from production.

Let us, to begin with, take relationship strength to be exogenous and symmetric across the supply network and examine aggregate output as we vary this parameter.

[2] Kremer (1993) is a seminal study of some theoretical aspects of such propagation. Carvalho et al. (2021) empirically study how shocks caused by the Great East Japan Earthquake of 2011 propagated through supply networks to locations far from the initial disruption. Barrot and Sauvagnat (2016) show that firms are reliant on their specific suppliers in the medium run.

[3] Though the basic shocks are modeled as independent, the interdependence between firms and their suppliers makes failures correlated between firms that (directly or indirectly) transact with each other.

Even this basic question about the mechanics of a supply network is not well understood. The key parameters, other than relationship strength, determining aggregate output are (i) the depth of supply chains, i.e., the number of steps of specialized sourcing; (ii) the number of distinct inputs required in each production process; (iii) the number of potential suppliers of each input. The first two dimensions capture the complexity of production, while the third captures the availability of multisourcing. In our model, there is a continuum of firms and the fraction of firms functioning is deterministic.[4] This fraction is our main outcome of interest, and we call it the *reliability* of the supply network. We are interested in the function mapping relationship strength to reliability. Our first main result concerns a distinctive form of sensitivity—what we call a precipice—in this function. Suppose production is complex in the sense that it is many steps deep, with multiple inputs at each step. There is an arbitrarily steep change in reliability as we vary relationship strength, holding all else fixed, when relationship strength transitions past a certain threshold (defining the precipice). Thus, if relationship strengths happen to be close to the precipice, a small, systemic, negative shock to relationship strengths is amplified arbitrarily strongly, leading to severe economic damage. We also show that a social planner choosing a level of investments in relationship strengths to maximize welfare would never choose a level such that the supply network is on a precipice.

Systemic shocks to relationship strengths can occur through several possible channels. Suppose, first, that the institutions that help uphold contracts and facilitate business transactions suddenly decline in quality, for example due to a political shock. Each supply relationship then becomes more prone to idiosyncratic disruptions.[5] Even if the damage to any single relationship is small (e.g., because usually contracts function without enforcement being relevant), our results show that such a shock can cause widespread disruptions throughout the supply network. Consider, next, a small shock to the availability of credit for businesses in the supply network. The shock matters for firms that are marginal recipients of loans essential for them to deliver on a commitment, and who do not receive such loans after the shock. The effect of such a credit shock can be modeled as any given supply relationship being slightly less likely to function (depending on ex ante uncertain realizations determining whether a firm is on the relevant margin). Third, during the COVID-19 crisis, technological disruptions and congestion delays degraded many relationships, with considerable uncertainty over which supply links would be affected, and at which times. This can be modeled as a systemic decrease in the probability that suppliers, in a given period, are able to deliver the inputs required from them.[6]

Our first result about the precipice took relationship strengths to be exogenous and showed that, starting at certain strength levels, supply network functioning *can* be very sensitive to slight shocks to strengths. However, relationship strengths are partly determined by strategic decisions. Our main results concern whether a supply network will be near a precipice when relationship strengths are shaped by equilibrium choices rather than being set exogenously or by a planner. The endogenous determination of relationship strengths is realistic. Since production is risky, an optimizing firm will strategically invest in relationships to manage the risk of its production being disrupted. By investing more, a firm can increase the probability that one of its potential suppliers for each of its essential inputs is able to supply it, hence allowing the firm to produce its output and make profits.[7] This investment is costly, and firms trade off these costs against the benefits of increased robustness.[8]

[4] This is by a standard diversification argument. There are enough supply chains that none of them is systemically important. On the complementary issue of systemically important firms, see Gabaix (2011) and the ensuing literature.

[5] Blanchard and Kremer (1997) present evidence that the former Soviet Union suffered a large shock of this kind when it transitioned to a market-based economy.

[6] Elliott and Golub (forthcoming) survey some facts on supply chain disruptions and their consequences.

Our main findings give conditions for equilibrium relationship strength levels to put the supply network on the precipice, and show that the precipice is *not* a knife-edge outcome. Indeed, we characterize a positive-measure set of parameters (governing the profits of production and the costs of forming relationships) for which the equilibrium supply network is on the precipice. The fragility that a supply network experiences in this regime is highly inefficient: a social planner would never put a supply network on the precipice for the same parameters. As supply networks become large and decentralized, one might think that the impact of uncertainty on the probability of successful production would be smoothed by firms' endogenous investments to protect against shocks and by averaging outcomes across a continuum population. We find the opposite: in equilibrium, there are powerful forces driving a supply network to a situation where aggregate productivity is very sharply sensitive to relationship strength. This is in contrast to many standard production network models, where the aggregate production function has bounded sensitivity to small shocks, even aggregate ones. The location of equilibrium on the precipice arises from three features of our environment. The first is complexity: The supply networks we study are many layers deep and firms must source multiple essential inputs that cannot be purchased off-the-shelf. The second is the presence of idiosyncratic disruptions that disable some links. The third is endogenous underinvestment in relationships.

In our main results, we vary an aggregate productivity parameter capturing the value of output relative to the costs of maintaining relationships, and examine whether the supply network is on a precipice. Depending on the value of this parameter, the supply network in equilibrium can end up in one of three configurations: (i) a *noncritical equilibrium* where the equilibrium investment is enough to keep relationship strength away from the precipice, (ii) a *critical equilibrium* where equilibrium relationship strength is on the precipice, and (iii) an *unproductive equilibrium* where positive investment cannot be sustained. These regimes are ordered. As the productivity of the supply network decreases from a high to a low level, the regimes occur in the order just given. Each regime occurs for a positive interval of values of the parameter. Equivalently, for an economy consisting of many disjoint supply networks distributed with full support over the parameter space, a positive measure of them will be in the fragile regime, and these networks will collapse if relationship quality is shocked throughout the economy.

[7] This can be interpreted in two ways: (i) investment on the intensive margin, e.g., to anticipate and counteract risks or improve contracts; (ii) on the extensive margin, to find more partners out of a set of potential ones.

[8] Strategic responses to risk in networks is a topic that has attracted considerable attention recently. See, for instance, Bimpikis, Candogan, and Ehsani (2019); Blume et al. (2011); Talamàs and Vohra (2020); Erol and Vohra (2018); Amelkin and Vohra (2019); and Acemoglu and Tahbaz-Salehi (2020). On the practical importance of the strength of contracts in supply relationships, see, among others, Antràs (2005); and Acemoglu, Antràs, and Helpman (2007).

Our analysis makes a conceptual, modeling, and technical contribution to the theory of economic networks. First, we introduce percolation analysis (i.e., disabling some links at random) to an otherwise standard network model of complex production—with *complex* meaning that each firm must source multiple inputs through customized relationships.[9] As a result, the aggregate production function exhibits a *discontinuous* phase transition, where aggregate functionality abruptly disappears when relationship strengths cross a critical threshold. *Simple* production, where each firm relies on risky sourcing for at most one type of input, is not susceptible to this fragility. Second, as a modeling contribution, we demonstrate the tractability of studying equilibrium investments in links—or, more precisely, investments in the probability that links are operational. Thanks to a continuous investment choice and a continuum of nodes, investment problems are characterized by relatively tractable first-order conditions, where firms take expectations over the randomness in network realizations. We expect the modeling devices we develop to have other applications. Finally, using our equilibrium conditions to deduce the ordering of regimes discussed above requires developing some new techniques for the analysis of large network formation games.[10] For example, a crucial step in our main results depends on showing that firms' investments in network formation are locally strategic substitutes at the equilibrium investment levels. This depends on subtle properties of equilibrium network structure and incentives, which we characterize.

We conclude by exploring some extensions and implications of our modeling. First, we examine robustness on a number of dimensions—e.g., in how we model shocks and relationship investments. Perhaps our most important robustness check relaxes certain symmetry or homogeneity assumptions imposed on the supply network for simplicity in our main analysis. We study how fragility manifests with firm heterogeneity. Precipices continue to obtain in the presence of rich heterogeneity across multiple dimensions: number of inputs required, multisourcing possibilities, directed multisourcing efforts, profitability, etc. One important additional implication of our heterogeneity analysis is that a supply network is only as strong as its weakest links: as one product enters the fragile regime, all products that depend on it directly or indirectly are simultaneously pushed into the fragile regime. Second, we interpret our main results in terms of the short- and medium-run resilience of a supply network to shocks, and consider whether the fragility we identify can be ameliorated by natural policy interventions. Third, we show how the supply networks we have studied can be embedded in a larger economy with intersectoral linkages that do not rely on specific sourcing. Our model yields a new channel for the propagation of shocks across sectors, and the stark amplification of these shocks. Fourth, while the focus of our analysis is on linking complex supply networks to aggregate volatility, we also discuss how the model can provide a perspective on some stylized facts concerning industrial development. After presenting our results, in Sections V and VI we discuss in detail how these results fit into the most closely related literatures.

[9] A recent model motivated by some of the same questions is Acemoglu and Tahbaz-Salehi (2020), where each node in a production network relies on one failure-prone custom supplier. In Section VI, we discuss related models on information sharing, financial contagion, and other settings.

[10] Because our networks are sparse, the methods we use differ from and complement the use of graphons (see Erol, Parise, and Teytelboym 2020).

## I. Model: The Supply Network

Our main object of study is a network whose nodes are a continuum of small firms producing differentiated products. These firms are connected to their suppliers by a network of potential supply relationships, a random subset of which are realized as operational for sourcing.

An outcome of central interest is the set of firms that are *functional*—i.e., capable of producing given the realization of operational relationships. This set is key in determining welfare and network formation incentives, which we introduce in Section III. The purpose of this section is to set up the basic structure of the production network and define the set of functional firms.

### A. *Nodes: Varieties of Products*

There is a finite set $\mathcal{I}$ of *products*. For each product $i \in \mathcal{I}$, there is a continuum $\mathcal{V}_i$ of *varieties* of $i$, with a typical variety $v$ being an ordered pair $v = (i,f)$, where $f \in \mathcal{F}_i \subseteq \mathbb{R}$ is a variety index. Each variety is associated with a small firm producing it. We take $\mathcal{F}_i = \left[0, \frac{1}{|\mathcal{I}|}\right]$ for all $i$, so that the total mass of varieties (and firms) in the supply network is 1. Let $\mathcal{V} = \bigcup_{i \in \mathcal{I}} \mathcal{V}_i$ be the union of all the varieties. These are the nodes in our supply network.

### B. *Links: Potential and Realized Supply Relationships*

First, for each product $i \in \mathcal{I}$ there is a set of required inputs $I(i) \subseteq \mathcal{I}$. Second, each variety $v \in \mathcal{V}$ is associated with a supply chain *depth* $d(v) \in \mathbb{Z}_+$ that specifies how many steps of customized, specifically sourced production are required to produce $v$, with varieties of larger depth requiring more steps. Different varieties of the same product can have different depths. The measure of varieties with any depth $d \geq 0$ is denoted by $\mu(d)$.

Consider any variety $v \in \mathcal{V}_i$. For each $j \in I(i)$ (i.e., each required input) the variety $v$ has a set of *potential suppliers* $\mathrm{PS}_j(v) \subseteq \mathcal{V}_j$, and a random subset of *realized suppliers* $\mathrm{S}_j(v) \subseteq \mathrm{PS}_j(v)$.

First consider the varieties $v \in \mathcal{V}$ such that $d(v) = 0$. Specialized sourcing of inputs is not required for these varieties, and so they operate without disruption. Thus, in this case, we take $\mathrm{S}_j(v) = \mathcal{V}_j$ for each $j \in I(i)$.

Next, consider any variety $v \in \mathcal{V}_i$ that has depth $d(v) > 0$. For each $j \in I(i)$, the set $\mathrm{PS}_j(v)$ is a finite set of distinct varieties $v' \in \mathcal{V}_j$ with each such $v'$ having depth $d(v') = d(v) - 1$. The identities of these suppliers are independent draws from the set of varieties $v'$ such that $d(v') = d(v) - 1$ (i.e., the set of varieties of compatible depth).[11] Specialized sourcing requirements represent the need for a customized input, the procurement of which is facilitated by relational contracts.

Each sourcing relationship between $v$ and a variety $v' \in \mathrm{PS}_j(v)$ is *operational* or not—a binary random outcome. For every $v \in \mathcal{V}$, there is a parameter $x_v$, called

[11] For a formal construction of the potential and realized supply networks, see online Appendix SA1.

*relationship strength* (for now exogenous), which is the probability that any relationship $v$ has with one of its potential suppliers is operational. All realizations of relationship operation are independent. The set of actual suppliers $\mathrm{S}_j(v)$ is then obtained by including each potential supplier in $\mathrm{PS}_j(v)$ independently with probability $x_v$. Whereas the potential supply relationships define compatibilities, the realized supply network identifies which links are actually available for sourcing. The stochastic nature of availability arises, e.g., from uncertainty in delivery of orders, miscommunications about specifications, etc.[12]

We define two random networks on the set $\mathcal{V}$ of nodes. In the *potential supply network* $\mathcal{G}$, each $v$ has links directed to all its potential suppliers $v' \in \bigcup_{j \in I(i)} \mathrm{PS}_j(v)$. (See Figure 1 for an illustration.) In the *realized supply network* $\mathcal{G}'$, each $v$ has only the subset of operational links to the realized suppliers: $v' \in \bigcup_{j \in I(i)} \mathrm{S}_j(v)$. See the links in Figure 2 for an illustration of the subset of supply relationships that are operational.

### C. *The Set of Functional Varieties*

For a given realization of the supply network, we will inductively define which varieties are *functional*, which means that sourcing disruptions do not prevent them from producing.

Depth-0 varieties are defined to be functional always: sourcing failures can never prevent the production of such varieties simply because their sourcing is unconstrained by definition. Given functionalities of varieties of depth $d - 1$, a variety $v$ of depth $d$ is functional if and only if its set of realized suppliers $S_j(v)$ contains at least one functional supplier for each input $j$ that $v$ requires. Figure 2 provides an illustration of the potential supply network shown in Figure 1 and a particular realization of operational supply links. It also illustrates the determination of which varieties are functional.

We let $\mathcal{V}'$ denote the (random) set of functional varieties, and $\mathcal{V}_i'$ denote the set of functional varieties in product $i$.

## II. Reliability with Exogenous Relationship Strengths

This section studies the size of the set of functional varieties when the distribution of shocks is given exogenously. It presents a key phenomenon in the mechanics of production: a discontinuous drop in aggregate production possibilities as relationship strength degrades.

### A. *Setup: Reliability in Regular Supply Networks*

We first present our main findings in *regular* supply networks. These are defined by two main symmetries. First, the number of essential inputs is $|I(i)| = m$ for

[12] In a bit more detail, $x_v$ can capture uncertainty regarding whether delivery can happen on time, possible misunderstandings about the required input, access to credit that may be needed to deal with unexpected costs, etc. It will depend on the context or environment in which production occurs, and also (as we explicitly model below) on the investments the firm producing $v$ makes. See Section V for more on interpretation.

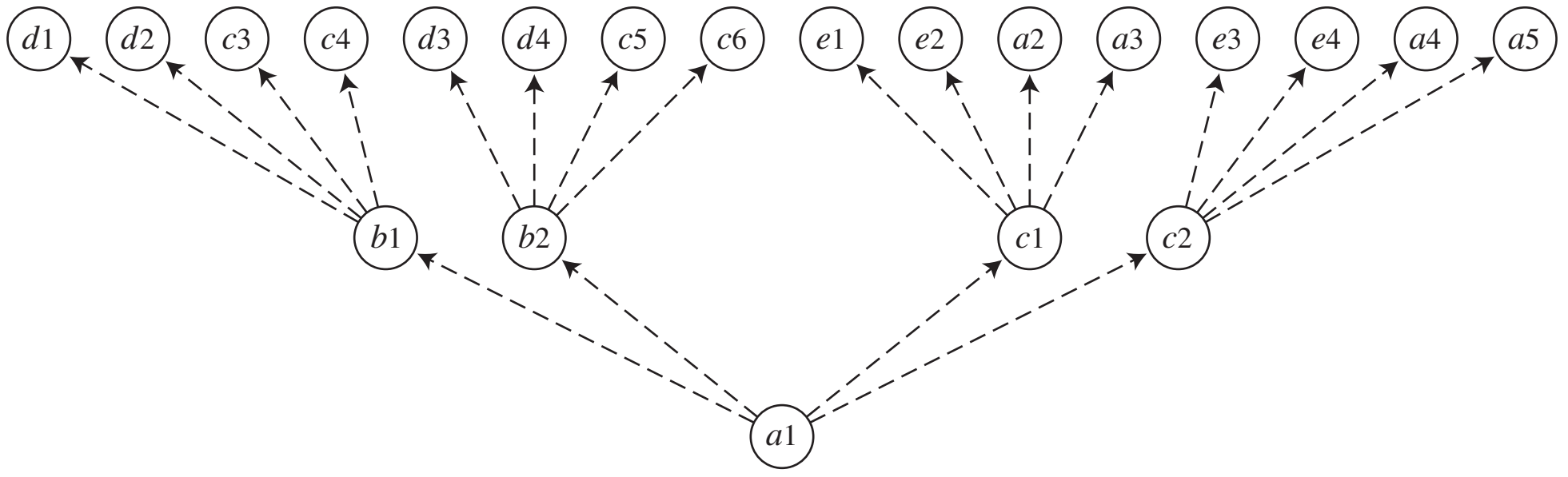

Figure 1

*Notes:* This illustration depicts the network of potential supply relationships upstream of a variety $a1$, in a supply network with underlying products $\mathcal{I} = \{a,b,c,d,e\}$. (We abbreviate $(a,1)$ as $a1$, and similarly for other varieties.) Each variety requires two distinct inputs; the relevant input requirements are apparent from the illustration. Here variety $a1$ has depth $d(a1) = 2$. Varieties higher up are upstream of $a1$, and their depths are smaller. For inputs that must be specifically sourced (i.e., those at a depth of 1 or greater), there is an edge from the sourcing variety to each potential supplier of the input. Orders or sourcing attempts go in the direction of the arrows, and products are delivered in the opposite direction, downstream.

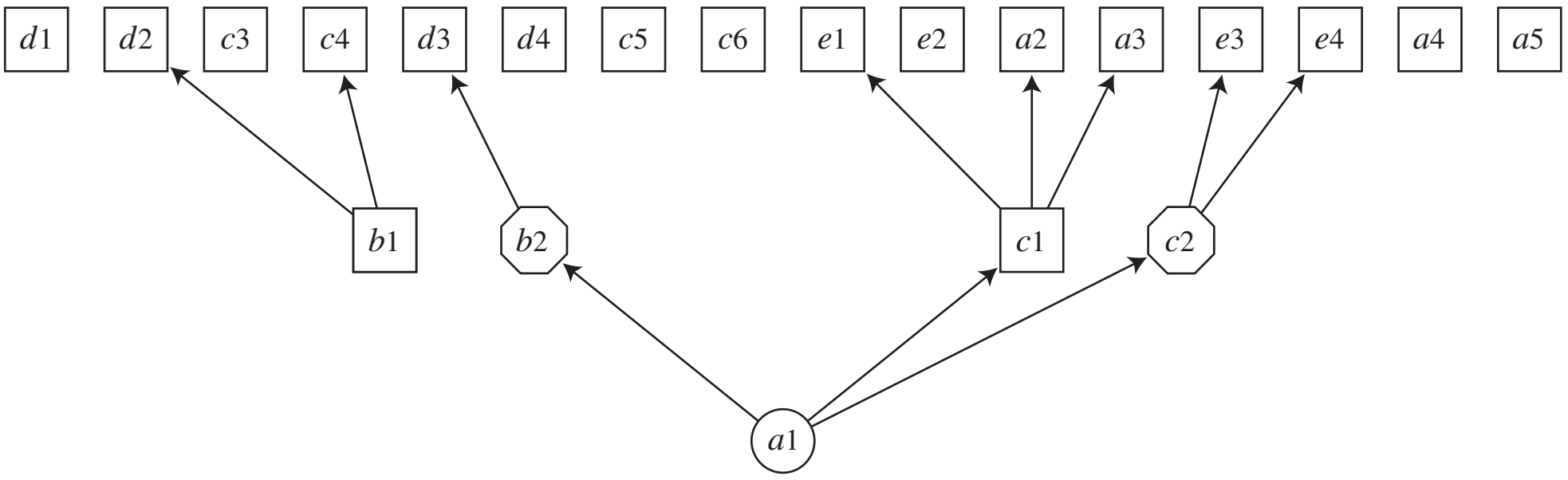

Figure 2

*Notes:* An illustration of the algorithm for determining the set of functional varieties given a realized supply network. Functional varieties are represented by squares, while nonfunctional ones are octagons. Varieties that have not yet been assigned a functionality status are circles. Varieties of depth 0 are always functional. This figure illustrates the step of the algorithm that assigns functionalities to varieties of depth 1. At this stage of the algorithm, no status is assigned to $a1$, the sole variety of depth 2. At the next step of the algorithm, $a1$ would become nonfunctional, changing into an octagon, since it has no functional supplier of the $b$ input.

each product $i$. Second, each variety of depth $d > 0$ has $n$ potential suppliers for each input—i.e., $|\mathrm{PS}_j(v)| = n$ whenever $j$ is one of the required inputs for variety $v$. Networks with these homogeneous structures are depicted in Figures 1 and 2. In this section, we posit that relationship strengths are given by $x_v = x$, the same number for each $v$, and this $x$ is exogenous.[13]

It will be important to characterize how the measure of functional firms $\mathcal{V}'$ varies with relationship strength $x$. Denote by $\tilde{\rho}(x,d)$ the probability that an arbitrary variety of depth $d$ is functional when relationship strength is $x$. Depth-0 varieties are functional by our assumption that they do not need any specialized inputs: $\tilde{\rho}(x,0) = 1$. By symmetry of the supply tree, the probability $\tilde{\rho}(x,d)$ does indeed only depend on $x$ and $d$ (and we calculate it explicitly below in Section IIC). The probability that a variety selected uniformly at random is functional is called the *reliability* of the supply network. Since it depends on the distribution of depths $\mu$, we denote this by $\rho(x,\mu)$ and define it as

$$\rho(x,\mu) = \sum_{d=0}^{\infty} \mu(d)\tilde{\rho}(x,d). \tag{1}$$

[13] We endogenize relationship strengths in the next section, and we relax the symmetry assumptions in Section VD.

*Deep Supply Networks: Taking Limits*.—A focus throughout will be the case where a typical variety has large depth.[14] We thus introduce a notation for asymptotics: we fix a sequence $(\mu_\tau)_{\tau=1}^{\infty}$ of distributions; $\tau$ is a parameter we will take to be large. We assume $\mu_\tau$ places probability at least $1 - \frac{1}{\tau}$ on $[D(\tau),\infty)$ for some unbounded, increasing function $D$.[15] For instance, we can take $\mu_\tau$ to be the geometric distribution with mean $\tau$ (in which case $\tau$ has an exact interpretation as average depth). For large $\tau$, if inputs are single-sourced ($n = 1$) and links fail with positive probability, there will only be a very remote probability of successful production. We therefore restrict attention to the case of multisourcing ($n \geq 2$).

## B. *A Discontinuity in Reliability*

A key implication of the model is the shape of the aggregate reliability function as we vary $x$. As we will see in the next section, this function is closely linked to the aggregate production function of the supply network, and thus its shape underlies many of our results. Our first result characterizes important properties of this shape.

PROPOSITION 1: *Fix any $n \geq 2$ and $m \geq 2$. Then there exist positive numbers $x_{\text{crit}}, \bar{r}_{\text{crit}} > 0$ such that,*

(*i*) *if $x < x_{\text{crit}}$, we have that $\rho(x,\mu_\tau) \to 0$ as $\tau \to \infty$. That is, reliability converges to 0 for relationship strength $x$ below $x_{\text{crit}}$.*

(*ii*) *if $x > x_{\text{crit}}$, then, for all large enough $\tau$, we have $\rho(x,\mu_\tau) > \bar{r}_{\text{crit}}$. That is, reliability remains bounded away from 0 for relationship strength $x$ above $x_{\text{crit}}$.*

In Figure 3 panel A we plot the reliability function $\rho(x,\mu_\tau)$ for a fixed finite value of $\tau$ against the probability $x$ of each relationship being operational. One can see a steep transition in relationship strength $x$. This can be seen more sharply in Figure 3 panel B, where we plot the limit of the graph shown in panel A as $\tau \to \infty$. We use the $m = 2, n = 4$ case here as in our illustrations above. There is a critical value of relationship strength, which we call $x_{\text{crit}}$, such that the probability of successful production is 0 when $x < x_{\text{crit}}$, but then increases sharply to more than 70 percent for all $x > x_{\text{crit}}$. Moreover, the derivative of the limit reliability graph as we

[14] Online Appendix SA7 investigates how reliability varies with investment in production trees with bounded depth.

[15] Note that this means varieties of low depth have many incoming edges in the potential supply network, since there are relatively few of them, but they make up a relatively large number of the nodes in a typical production tree.

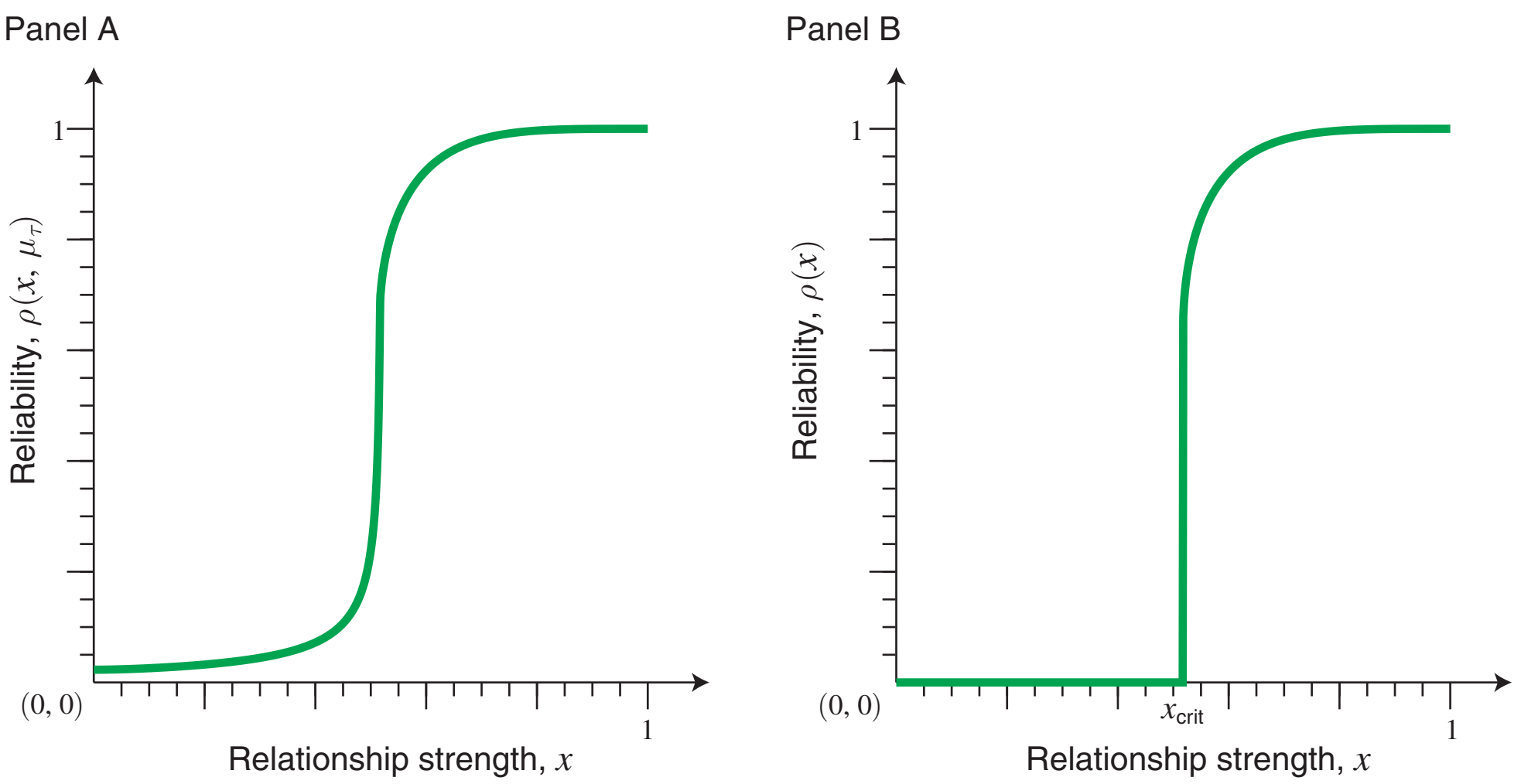

Figure 3

*Notes:* Panel A shows how reliability varies with relationship strength $x$ for a particular $\tau$. Panel B depicts a correspondence that is the limit of the graphs $\rho(x, \mu_\tau)$ as $\tau$ tends to infinity.

approach $x_{\text{crit}}$ from above grows arbitrarily large (i.e., $\lim_{x \downarrow x_{\text{crit}}} \lim_{\tau \to \infty} \rho'(x, \mu_\tau) = \infty$). This has important ramifications, as we will see. An immediate one is that small improvements in relationship strength $x$, for example through the improvement of institutions, can have large payoffs for an economy, and the net marginal returns on investment in $x$ can change sharply from being negative to being positive and very large.

### C. *The Reasons for the Shape of the Reliability Function*

To explain the logic behind the proposition, let us now calculate the probability that a given variety $v$ with depth $d$ is functional. Recall that we denote by $\tilde{\rho}(x, d)$ the probability that a variety of depth $d$ is functional when relationship strength is $x$. We will argue that this can be expressed recursively as follows. First, $\tilde{\rho}(x, 0) = 1$, since varieties of depth 0 are sure to be functional. Then, for a suitably defined function $\mathcal{R}_x : [0,1] \to [0,1]$, we can write the depth-$d$ reliability in terms of the depth-$(d-1)$ reliability:

$$\tilde{\rho}(x, d) = \mathcal{R}_x\big(\tilde{\rho}(x; d-1)\big). \tag{2}$$

Indeed, more explicitly, the function that makes this true is[16]

$$\mathcal{R}_x(r) = \big(1 - (1 - xr)^n\big)^m.$$

[16] Fix a variety and consider any one of its inputs. For a given supplier of that input, by definition its reliability is the argument $r$, and the probability that the link to the supplier is operational is $x$. The probability that both events happen is $xr$. The probability that this combination of events happens for at least one of the $n$ potential suppliers of the first input is therefore $1 - (1 - xr)^n$. Finally, the probability that for all $m$ inputs, such a "good event" happens is $(1 - (1 - xr)^n)^m$.

As we look at networks with large typical depths, it is the $\tilde{\rho}(x,d)$ for large $d$ that will matter. Proposition SA1 in online Appendix SA2.1 shows that there is a unique correspondence $\rho(x)$ that is the limit of the graphs of $\rho(x,\mu_\tau)$ in a suitable sense as $\tau \to \infty$. By analyzing this correspondence we can show that a sharp transition like that shown in Figure 3 panel B occurs for any complexity $m \geq 2$ and any multisourcing level $n \geq 2$, once depths become large.

The intuition for the sharp transition is illustrated in Figure 4. In panel A, we plot the probability that a given firm is functional against the probability that its suppliers are functional (which is taken to be common across the suppliers, and denoted by $r$). The curve $\mathcal{R}_x(r)$ is shown for several values of $x$. The shape of this curve, and how it moves as we vary $x$, will be crucial to explaining the shape of the reliability curve. We first examine the determination of equilibrium reliability given one of these curves. In Figure 5, we zoom in on the top-right corner of Figure 4 panel A—note the change in axes—and sketch one of these curves for a particular value of $x$, namely $x = 0.55$, along with the 45-degree line. We start at $r = \tilde{\rho}(x,0) = 1$ and apply $\mathcal{R}_x$ to this value; then we apply $\mathcal{R}_x$ to the output of that, and so on. Figure 5 shows how, in doing so, we obtain a sequence of $\tilde{\rho}(x,d)$ converging to the largest fixed point of $\mathcal{R}_x(r)$. As supply networks become deep, a firm and its suppliers occupy essentially equivalent positions, so it is natural that the limit probability of being functional for large-depth firms is equal to such a fixed point.

Returning to Figure 4, the reliability levels $r$ that are fixed points of the function $\mathcal{R}_x(r)$ in equation (2) are given by its intersections with the 45-degree line, illustrated in panel A for several values of $x$ (each corresponding to one of the curves). We will now examine how the largest such intersection depends on $x$. For high enough $x$, the function $\mathcal{R}_x(r)$ has a fixed point with $r > 0$. When $x$ is below a certain critical value $x_{\text{crit}}$, the graph of $\mathcal{R}_x$ has an intersection with the 45-degree line only at $(0,0)$, and so the limit of $\tilde{\rho}(x,d)$ as $d \to \infty$ is 0. Crucially, the largest fixed point of $\mathcal{R}_x(r)$ does not decrease continuously to 0 as we lower $x$. Instead, it drops down discontinuously when $x$ decreases past $x_{\text{crit}}$—defined as the (positive) value of $x$ where there is a point of tangency between $\mathcal{R}_x(r)$ and the 45-degree line. At this point, the largest $r$ solving $r = \mathcal{R}_x(r)$ jumps down discontinuously from the $r$ corresponding to this point of tangency to 0. This is the discontinuous "precipice" drop. As we explain in Section VB, where we contrast complex production $(m > 1)$ with simple production $(m = 1)$, the convex-then-concave shape of the $\mathcal{R}_x$ curve is essential for creating a precipice.

*Comparative Statics of the Reliability Function.*—Some straightforward comparative statics can be deduced from what we have said. If $n$ (multisourcing) increases while all other parameters are held fixed, then one can check that $\mathcal{R}_x$ (as illustrated in Figure 4 panel A) increases pointwise on $(0,1)$, and this implies that all the $\tilde{\rho}(x,d)$ increase. It follows that the $\rho$ curve moves upward, and the discontinuity occurs at a lower value of $x$.

Similarly, when $m$ (complexity) increases, the $\mathcal{R}_x$ curve decreases pointwise, implying that all the $\tilde{\rho}(x,d)$ decrease. It follows that the $\rho$ curve moves downward, and the discontinuity occurs at a higher value of $x$.

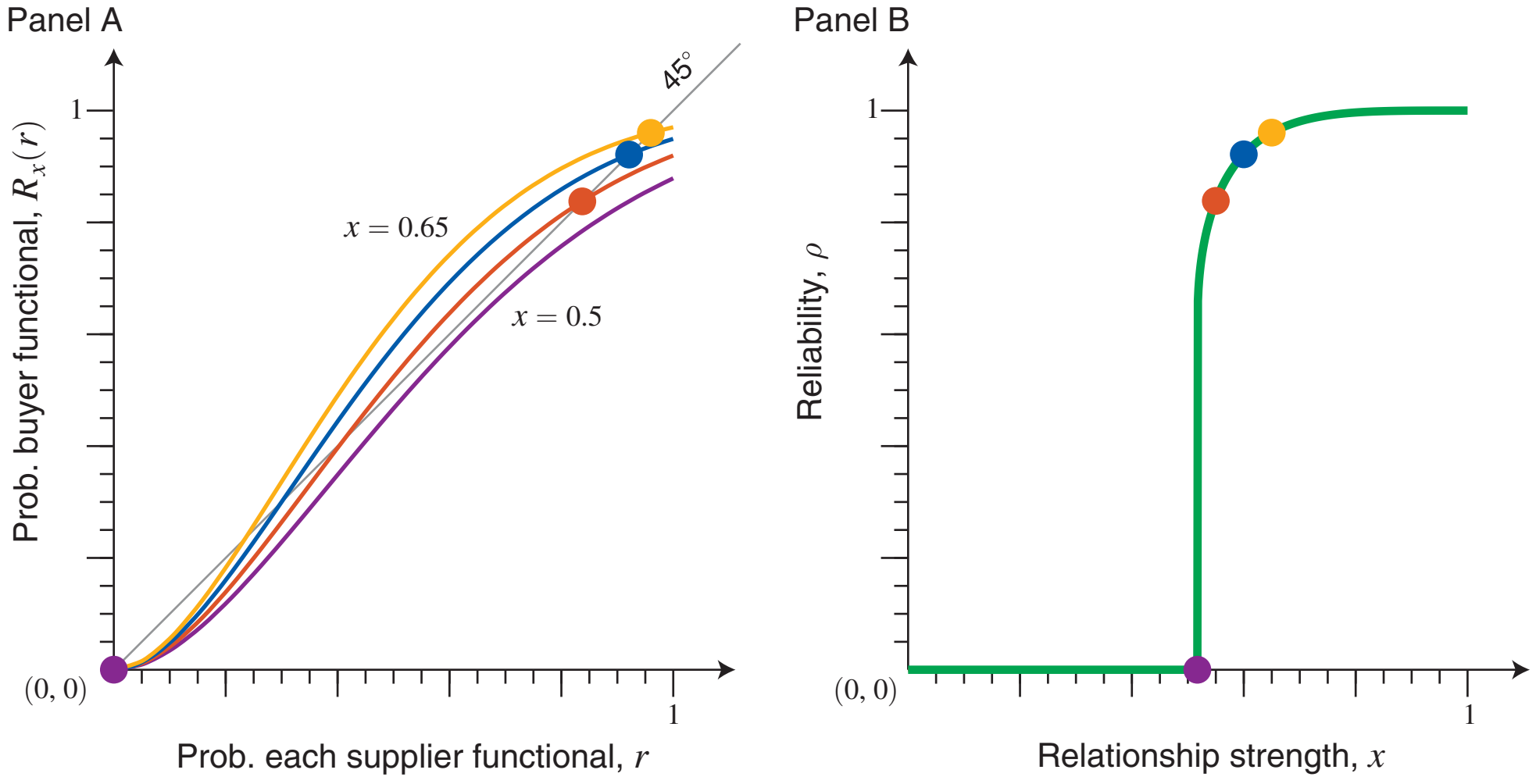

Figure 4

*Notes:* Panel A shows the probability, $\mathcal{R}_x(r)$, that a focal firm is functional as a function of $r$, the probability that a random supplier is functional. Here we use the parameters $n = 4, m = 2$, and $x \in \{0.5, 0.55, 0.6, 0.65\}$. The intersections with the 45-degree line marked by the circles represent reliability values (for deep supply trees) associated with the given exogenous parameters. Panel B shows the limit reliability, $\rho(x)$, as a function of $x$. Note that, for each $x$, the value of $\rho(x)$ is at the same height as the intersection of the corresponding $\mathcal{R}_x$ curve with the 45-degree line in panel A.

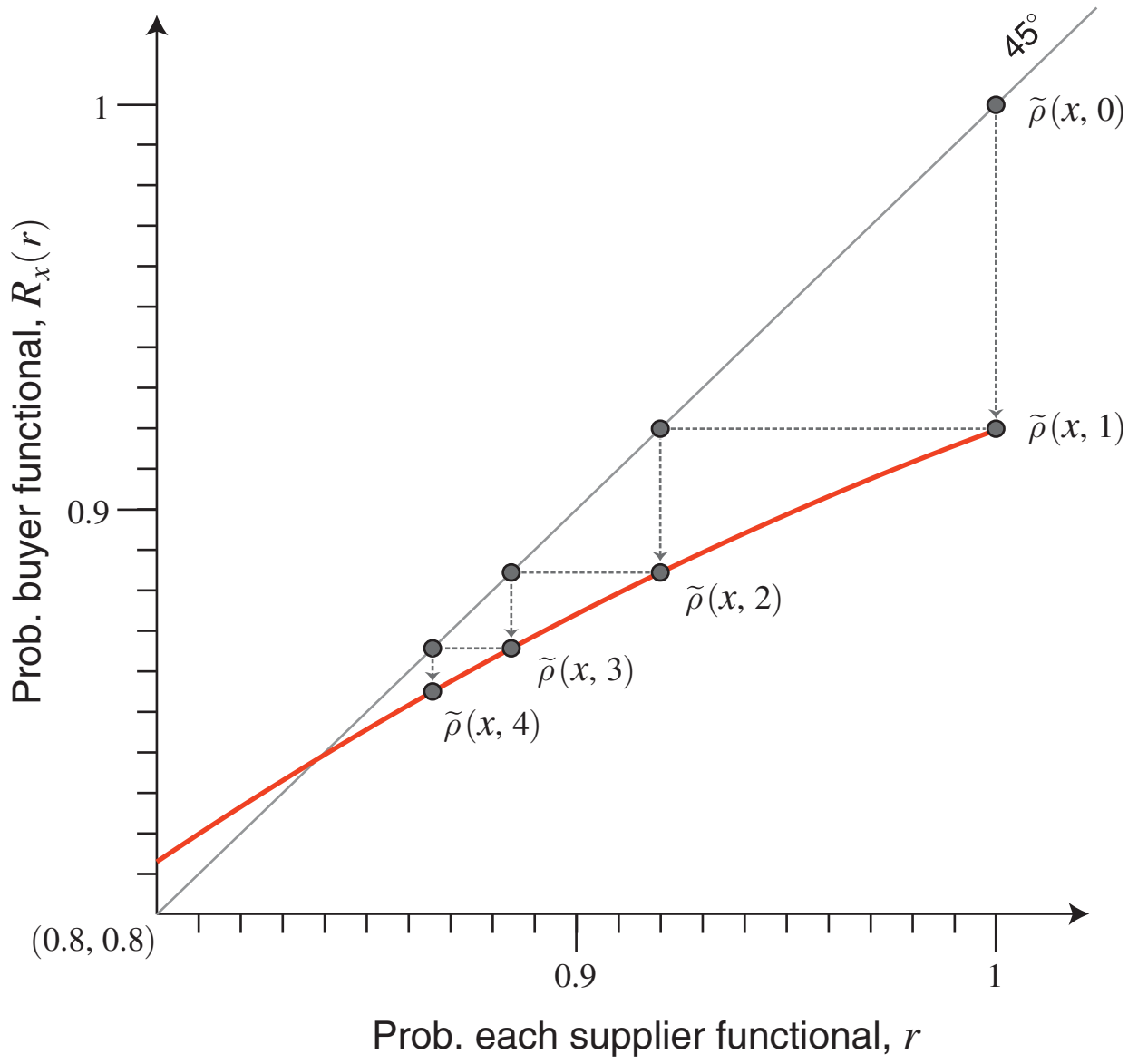

Figure 5

*Notes:* The curve $\mathcal{R}_x(r)$ is shown for values of $r \in [0.8, 1]$, for relationship strength $x = 0.55$. The values of $\tilde{\rho}(x, d)$ for $d \in \{0, 1, \ldots, 4\}$ given by equation (2) are depicted. Note that the largest intersection of $\mathcal{R}_x(r)$ with the 45-degree line is approached quite closely after a few steps.

## III. Supply Networks with Endogenous Relationship Strength

In this section, we study the endogenous determination of relationship strength. We first examine a planner's problem, in which strength is chosen in a centralized way, and show that investments that put the supply network on the precipice are never chosen by a planner. We then set up a decentralized problem in which firms invest in their own relationship strengths. Throughout, we focus on symmetric outcomes.

To connect the graph-theoretic notion of reliability in the previous section with economic outcomes, we introduce a relationship between reliability and output. In particular, we posit that there is a function $Y$ such that $Y(\mathcal{V}')$ is the (expected) aggregate gross output associated with a given set $\mathcal{V}'$ of functional firms. We assume that it has the following property:

**Property A:** In a regular supply network, gross output satisfies $Y(\mathcal{V}') = h(\rho(x,\mu))$, where $h : [0,1] \to \mathbb{R}$ is a strictly increasing, concave function with bounded and continuous derivative. Moreover, $h(0) = 0$.

In online Appendix SA3, we provide a microfoundation for this assumption: we define a standard production network model on top of the supply network, being explicit about production functions and markets at the microlevel. There, we show that aggregate output satisfies Property A. However, our main results—about efficient and equilibrium networks—do not rely on any details of the microfoundations. Instead, they rely only on assumptions about benefits and costs of production that we bring out into named properties.

Property A sets the stage for our study of the planner's problem.

### A. *A Planner's Problem*

We study a planner who chooses a global $x$ that determines the values of all relationship strengths, $x_v = x$. This can be interpreted as investing in the quality of institutions, at a cost that we will introduce below. As stipulated in Property A, the gross output of the supply network is an increasing function of reliability, $Y(\mathcal{V}') = h(\rho(x,\mu))$, where reliability $\rho$ depends on the symmetric level of relationship strengths $x$ and the depth distribution $\mu$. The planner's cost of a given choice of $x$ enters through subtracting a quantity $\frac{1}{\kappa}c_P(x)$ of output, where $\kappa$ is a strictly positive *productivity parameter* and $c_P : [0,1] \to \mathbb{R}$ is a fixed function. Higher values of $\kappa$ have the interpretation that they shift down the costs of obtaining a given level of relationship strengths, i.e., of obtaining a given level of gross output. The planner seeks to maximize expected net aggregate output by choosing relationship strengths $x$, and hence solves the *planner's problem:*

$$(3) \qquad \max_{x\in[0,1]} h(\rho(x,\mu)) - \frac{1}{\kappa}c_P(x).$$

We make the following assumption concerning $c_P(x)$:

**Property B:** The planner's cost function $c_P : [0,1] \to \mathbb{R}$ is continuously differentiable and weakly convex, with $c_P(0) = 0$, $c_P(x_{\text{crit}}) > 0$, $c_P'(0) = 0$, and $\lim_{x\to 1} c_P'(x) = \infty$.

Substantively, the conditions in this property entail that there are increasing marginal social costs of relationship strength and achieving the critical level of reliability requires a positive investment. The Inada conditions on derivatives ensure good behavior of optima.

Define the correspondence

$$x^{\text{SP}}(\kappa,\mu) := \underset{x\in[0,1]}{\operatorname{argmax}}\left[h(\rho(x,\mu)) - \frac{1}{\kappa}c_P(x)\right].$$

This gives the values of $x$ that solve the social planner's problem for a given $\kappa$ and distribution of supply chain depths $\mu$. As elsewhere, we consider a sequence $(\mu_\tau)_{\tau=1}^{\infty}$ of depth distributions, where $\mu_\tau$ places mass at least $1 - \frac{1}{\tau}$ on $[D(\tau),\infty)$ for an unbounded, increasing function $D$.

PROPOSITION 2: *Fix any $n \geq 2$ and $m \geq 2$. Then there exists a number $\kappa_{\text{crit}} > 0$ such that, for all small enough $\epsilon > 0$, the following statements hold for all large enough $\tau$:*

(*i*) *For all $\kappa < \kappa_{\text{crit}}$, all values of $x^{\text{SP}}(\kappa,\mu_\tau)$ are at most $x_{\text{crit}} - \epsilon$, have cost less than $\epsilon$, and yield reliability less than $\epsilon$;*

(*ii*) *For all $\kappa > \kappa_{\text{crit}}$, all values of $x^{\text{SP}}(\kappa,\mu_\tau)$ are at least $x_{\text{crit}} + \epsilon$ and yield reliability at least $r_{\text{crit}} + \epsilon$;*

(*iii*) *For $\kappa = \kappa_{\text{crit}}$, all values of $x^{\text{SP}}(\kappa,\mu_\tau)$ are outside the interval $[x_{\text{crit}} - \epsilon, x_{\text{crit}} + \epsilon]$ and reliability is outside the interval $[\epsilon, r_{\text{crit}} + \epsilon]$.*

The first part of Proposition 2 says that when $\kappa$ is sufficiently low, it is too costly for the social planner to invest anything in the quality of institutions, and hence reliability is very low. As $\kappa$ crosses a threshold $\kappa_{\text{crit}}$, it first becomes optimal to invest in institutional quality. At this threshold, the social planner's investment increases discontinuously. Moreover, it immediately increases to a level strictly *above* $x_{\text{crit}}$, and, for all larger $\kappa$, all solutions stay above $x_{\text{crit}}$.

It is worth emphasizing that the planner never chooses to invest near the critical level $x_{\text{crit}}$. The reason is as follows: sufficiently close to $x = x_{\text{crit}}$, the marginal social benefits of investing grow arbitrarily large in the limit as $\tau$ gets large while marginal costs at $x_{\text{crit}}$ are bounded, and so the social planner can always do better by increasing investment at least a little. In contrast, in Section IV we will see that individual investment choices can put the supply network on the precipice in equilibrium, and this is not a knife-edge scenario.

## B. *Decentralized Investment in Relationship Strengths*

Now we formulate a simple, symmetric model of decentralized choices of relationship strengths.

*Setup and Timing*.—The decision-makers in this richer model are firms. For each product $i$, there is a continuum of separate firms $(i,f)$, where $f \in [0, 1/|\mathcal{I}|]$. The firm $(i,f)$ owns the corresponding variety, $v = (i,f)$; our notation identifies a firm with its variety. We often abbreviate both by $if$.

Firms simultaneously choose *investment levels* $y_{if} \geq 0$. Choosing a level $y_{if}$ has a private cost $c(y_{if})$. The random realization of the supply network occurs after the firm chooses its investment level.[17] If a firm chooses an investment level $y_{if}$, then all sourcing links from its variety $(i,f)$ have relationship strength

$$x_{if} = \underline{x} + y_{if}.$$

The intercept $\underline{x} \geq 0$ is a baseline relationship strength that obtains absent any costly investment. This can be interpreted as a measure of the quality of institutions—e.g., how likely a "basic contract" is to deliver.[18] The main purpose of this baseline level is as a simple channel to systemically shock relationship strengths throughout the supply network.

The timing is as follows:

1. Firms simultaneously choose their investment levels.

2. The realized supply network is drawn and payoffs are enjoyed.

An *outcome* is given by relationship strengths $x_{if}$ for all firms $if$. An outcome is symmetric if all firms have the same relationship strength: $x_{if} = x$ for all $if$.

*Payoffs and Equilibrium*.—A firm's payoff at an outcome can be written as

$$u_{if} = G_{if} - \frac{1}{\kappa} c(x_{if} - \underline{x}).$$

This is the firm's expected gross profit, $G_{if}$, minus the cost of its investment, $y_{if} = x_{if} - \underline{x}$, in relationship strength. We will now discuss the parts of this payoff function in turn.

We begin with the firm's costs. We assume that they satisfy the following property, where $\kappa > 0$ is a parameter:

**Property B′:** A firm's cost is given by $\frac{1}{\kappa} c(x_{if} - \underline{x})$, where the following conditions hold:

(i) $\underline{x} < x_{\text{crit}}$;

[17] The supply network realization is defined as an assignment of depths to all varieties, and the graphs $\mathcal{G}$ and $\mathcal{G}'$ from Section I. The assumption that investments are made before this realization is technically convenient, as it keeps the solution of the model symmetric. For example, a firm knows that after some number of stages of production, disruption-prone contracts will not be needed by its indirect suppliers (e.g., because these suppliers are able to use generic inputs or rely on inventories). However, the firm does not know how many steps this will take. See the first part of Section VD for an extension where firms have some information about their depths.

[18] A natural interpretation is that this is a feature of the contracting environment—concretely, for instance, it could reflect the quality of the commercial courts.

(ii) $c'$ is increasing, continuously differentiable, and strictly convex, with $c(0) = 0$;

(iii) the following limit conditions hold: $\lim_{y \downarrow 0} c'(y) = 0$ and $\lim_{y \uparrow 1-\underline{x}} c'(y) = \infty$.

The first part of this assumption ensures that baseline relationship strength is not so high that the supply network is guaranteed to be productive even without any investment. The second part imposes assumptions on investment costs that ensure agents' optimization problems are well-behaved. The limit conditions ensure that investments are interior. Here $\kappa$ plays the same role as it did in our social planner optimization exercise: scaling down the costs of investing in relationship strength—and hence achieving a given level of productivity.

We now turn to specifying gross profits at a given outcome. Because we will characterize symmetric equilibria, we need to specify the gross profits of a given firm only for symmetric behavior by other firms. Firms make no gross profits conditional on not producing, and their profits conditional on producing satisfy the following assumption.

**Property C:** At a symmetric outcome with reliability $r$, conditional on being functional, a firm makes gross profits $g(r)$, where $g : [0,1] \to \mathbb{R}_+$ is a decreasing, continuously differentiable function.

Property C requires that profits are higher when fewer firms are functioning and there is less competition. Online Appendix SA3 microfounds this property in the same production network model that we used to microfound Property A.

Let $P(x_{if}; x, \mu)$ be the probability that a firm $if$ is functional if firm $if$'s relationship strength is $x_{if}$ and all other firms choose symmetric relationship strengths $x$. Then, under Property C, we have that $G_{if} = P(x_{if}; x, \mu) g(r)$, where $r$ is the reliability associated with $x$. Thus, recalling the payoff formula given at the start of this section, the net expected profit of firm $if$ when it has relationship strength $x_{if}$ and all other firms have relationship strengths $x$, resulting in reliability $r$, is

$$\Pi_{if} = P(x_{if}; x, \mu) g(r) - \frac{1}{\kappa} c(x_{if} - \underline{x}). \tag{4}$$

Finally, because welfare properties of equilibria will play a role in our analysis, we define social welfare (for symmetric outcomes). The gross output of production given reliability $r$ has a value of $h(r)$, as in the previous section on the planner's problem. (Some of this goes to gross profits and some to consumer welfare, but the sum is given by $h(r)$.) The planner's cost function is simply the total of firms' costs,

$$\sum_{i \in \mathcal{I}} \int_{f \in \mathcal{F}_i} \frac{1}{\kappa} c(x - \underline{x}) \, df = \frac{1}{\kappa} c(x - \underline{x}),$$

where we have used our assumption that the mass of firms in each industry is $1/|\mathcal{I}|$, so that the total mass of firms is 1. Thus, if $x$ is the relationship strength of all firms, the social welfare function is[19]

$$\text{Welfare} \;=\; h\big(\rho(x,\mu)\big) - \frac{1}{\kappa}c(x - \underline{x}). \tag{5}$$

## IV. Equilibrium Supply Networks and Their Fragility

We now study equilibrium decentralized investment in relationship strengths in the model we have introduced. We are interested in the productivity and robustness of equilibrium supply networks. This section builds up to a main result: Theorem 1. We show that in the limit as production networks become deep, there are three regimes. First, for low values of the parameter $\kappa$, there is an unproductive regime in which equilibrium reliability is arbitrarily low. Next, for intermediate values of $\kappa$, there is a critical regime in which equilibrium relationship strengths are very close to $x_{\text{crit}}$ and arbitrarily small shocks to relationship strength lead to discontinuous drops in production. Finally, there is a noncritical regime in which equilibrium relationship strengths are above $x_{\text{crit}}$ and the supply network is robust to small shocks.

### A. *Equilibrium Reliability*

To examine firms' incentives, it is worth writing a firm's profit, (4), more explicitly. To do this, we calculate $P\big(x_{if}; x, \mu\big)$, the probability that $if$ is able to produce, as a function of firm $if$'s relationship strength $x_{if}$, given that all other firms choose a symmetric relationship strength $x$:[20]

$$P\big(x_{if}; x, \mu\big) \;:=\; \mu(0) + (1-\mu(0))\,\mathbb{E}\left[1 - \Big(1 - x_{if}\tilde{\rho}(x, d-1)\Big)^{n}\right]^{m},$$

where $d$ inside the expectation is drawn from the depth distribution $\mu$ conditional on depth being at least 1. (Recall from equation (2) in Section IIC the formula for $\tilde{\rho}(\,\cdot\,)$, the reliability of a variety of a given depth.)

DEFINITION 1: *We say $x \geq \underline{x}$ is a symmetric equilibrium if $x_{if} = x$ maximizes $\Pi_{if}\big(x_{if}; x, \mu\big)$ defined in (4) for any firm if. It is a symmetric undominated equilibrium if it is a symmetric equilibrium and, among symmetric equilibria, maximizes the social welfare defined in (5).*

When we refer to an equilibrium in the sequel, we mean a symmetric undominated equilibrium unless otherwise noted. Note that a symmetric equilibrium is defined by the level of relationship strength $x = \underline{x} + y$ realized in it, rather than the level of investment $y$. This turns out to be more convenient.

[19] This coincides with the welfare function of Section IIIA if we set $c_P(x) = c(x - \underline{x})$ for $x \geq \underline{x}$, and $c_P(x) = 0$ otherwise.

[20] Note that because there is a continuum of firms, the probability that a firm appears in its potential supply network upstream of itself is 0. Thus the reliability of $if$'s suppliers does not depend on $x_{if}$.

Our equilibrium definition requires that all firms' investment choices are equal and are mutual best responses to each other. Lemma SA4 in the online Appendix shows that the efficiency condition selects the symmetric equilibrium associated with the highest investment level, and hence highest reliability. Note that it is always a best response for a firm to choose zero investment when all others choose zero investment. Our equilibrium definition abstracts from potential miscoordination on the zero investment level, or other inefficient ones, by selecting the symmetric equilibrium that maximizes welfare.

In the limit, as the expected depth of the supply networks becomes large, if firms symmetrically choose investments $y_{if} = 0$, then the reliability is $\rho(\underline{x}) = 0$ as $\underline{x} < x_{\text{crit}}$ (by Property B$'$). Hence, for large enough $\tau$, $x_{if} = \underline{x}$ maximizes $\Pi_{if}(x_{if}; \underline{x}, \mu_\tau)$ and so there always exists an equilibrium.

In analyzing the symmetric equilibria it is helpful to make an assumption on the environment that ensures that the first-order conditions of firms' problems are sufficient for optimality among interior solutions. We first state the assumption and then formulate a condition on primitives that is sufficient for it to hold.

ASSUMPTION 1: *For any $\tau$ and any $x > x_{\text{crit}}$, the function $x_{if} \mapsto \Pi_{if}(x_{if}; x, \mu_\tau)$ has exactly one interior local maximum for all large enough $\tau$.*[21]

Assumption 1 will be maintained in the sequel, along with Properties A, B$'$, and C. The following lemma shows that we may always set $\underline{x}$ so that Assumption 1 is satisfied.

LEMMA 1: *For any $m \geq 2$ and any $n \geq 2$, there is a number*[22] *$\hat{x}$, depending only on $m$ and $n$, such that, for large enough $\tau$ we have (i) $\hat{x} < x_{\text{crit}}$, and (ii) if $\underline{x} \geq \hat{x}$, then Assumption 1 is satisfied.*

To see why this lemma implies that Assumption 1 is satisfied for a suitable choice of $\underline{x}$, consider any environment where $\underline{x} \in [\hat{x}, x_{\text{crit}})$. Part (i) of the lemma guarantees that the interval $[\hat{x}, x_{\text{crit}})$ is nonempty, and part (ii) guarantees that Assumption 1 is satisfied for values of $\underline{x}$ in this range.[23] In other words, the lemma guarantees that there is an interval of possible baseline levels of strengths that are short of the critical level (so that fragility is not ruled out a priori) but high enough to ensure that the firms' maximization problem is amenable to a first-order approach.

We now characterize the equilibrium behavior.

THEOREM 1: *Fix any $n \geq 2$, $m \geq 3$, and $\epsilon > 0$.*[24] *There are thresholds $\underline{\kappa}(n,m) < \bar{\kappa}(n,m)$ and a threshold $\underline{\tau}$, so that for all $\tau \geq \underline{\tau}$ there is a unique symmetric undominated equilibrium. The relationship strength in this equilibrium, denoted by $x_\tau^*(\kappa)$, satisfies the following properties:*

[21] The assumption permits another local maximum at a corner. We rule this out separately.

[22] In the proof, we give an explicit description of $\hat{x}$ in terms of the shape of the function $P(x_{if}; x, \mu_\tau)$.

[23] While conditions we identify in Lemma 1 are sufficient for satisfying Assumption 1, they are not necessary.

[24] In this result, we restrict attention to the case of $m \geq 3$. It is essential for our results that supply networks are complex ($m \geq 2$), but the case of $m = 2$ generates some technical difficulties for our proof technique so we consider $m \geq 3$. In numerical exercises, our conclusions seem to also hold for $m = 2$.

(*i*) *If* $\kappa < \underline{\kappa}$, *there is no investment*: $x_{\tau}^{*}(\kappa) = \underline{x}$.

(*ii*) *For* $\kappa \in [\underline{\kappa}, \overline{\kappa}]$, *the equilibrium relationship strength satisfies* $x_{\tau}^{*}(\kappa) \in [x_{\text{crit}} - \epsilon, x_{\text{crit}} + \epsilon]$. *We call such equilibria* ***critical***.

(*iii*) *For* $\kappa > \overline{\kappa}$, *the equilibrium relationship strength satisfies* $x_{\tau}^{*}(\kappa) > x_{\text{crit}}$. *We call such equilibria* ***noncritical***.

*Moreover, for* $\tau \geq \underline{\tau}$, *the function* $x_{\tau}^{*}(\kappa)$ *is increasing on the domain* $\kappa > \underline{\kappa}$.

If we think of different supply networks as being parameterized by different values of $\kappa$ in a finite interval, Theorem 1 implies that in the limit as $\tau$ gets large, there will be a positive fraction of supply networks in which firms will choose relationship strengths converging to $x_{\text{crit}}$ in equilibrium. This contrasts with the social planner's solution, which never selects relationship strengths near $x_{\text{crit}}$. It also means that a positive fraction of supply networks end up perched on the precipice, vulnerable to small unanticipated systemic shocks. This vulnerability extends to anticipated shocks, as discussed in Section VC.

*The Reasons behind the Theorem.*—Figure 6 helps give some intuition for Theorem 1. In any symmetric equilibrium, the reliability of each firm must be consistent with the symmetric investment level $x$ chosen by all other firms—we must be somewhere on the reliability curve we derived in Section IIB. The graphs labeled $\rho(x)$ in panels A–D of Figure 6 illustrate the shape of the reliability curve for large $\tau$. Further, in any symmetric equilibrium each firm's (symmetric) investment choice of $x$ must be a best response to the reliability of its suppliers. The curves labeled $\text{BR}(r)$ in panels A–D depict the best-response function; these curves should be thought of as having their independent variable (others' reliability) on the vertical axis and the best-response investment on the horizontal axis. Note that the entire curve $\text{BR}(r)$ depends on $\kappa$; the panels show the best-response curves for increasing values of $\kappa$. Intersections of these two curves correspond to potential symmetric equilibria. When there are multiple intersections, we focus on the one associated with the highest reliability, which is the one that is selected by Definition 1 (see Lemma SA4 in the online Appendix). Thus equilibrium reliability is 0 for $\kappa$ sufficiently small, jumps up discontinuously to $\underline{r}_{\text{crit}}$ at $\underline{\kappa}$, and is increasing in $\kappa$ thereafter. This is shown in panel E. Reliability clearly increases as $\kappa$ ranges over the interval $[\underline{\kappa}, \overline{\kappa}]$. Nevertheless, for $\kappa$ in this interval, equilibrium investment for large $\tau$ remains arbitrarily close to $x_{\text{crit}}$ since the equilibrium is on the essentially vertical part of the reliability curve. In other words, equilibrium investment choices bunch around $x_{\text{crit}}$ for an open interval of values of $\kappa$. For all $\kappa \in [\underline{\kappa}, \overline{\kappa}]$ a slight shock causing relationship strengths to diminish from $\underline{x}$ to $\underline{x} - \epsilon$ causes relationship strengths to fall below $x_{\text{crit}}$, and makes equilibrium production collapse. Panel F shows reliability after such a shock for different values of $\kappa$. As can be seen by comparing panels E and F, the shock does not change the values of reliability at $\kappa$ below $\underline{\kappa}$ or above $\overline{\kappa}$: the curves in E and F are essentially identical in those ranges. However, for $\kappa \in [\underline{\kappa}, \overline{\kappa}]$, reliability drops discontinuously to 0.

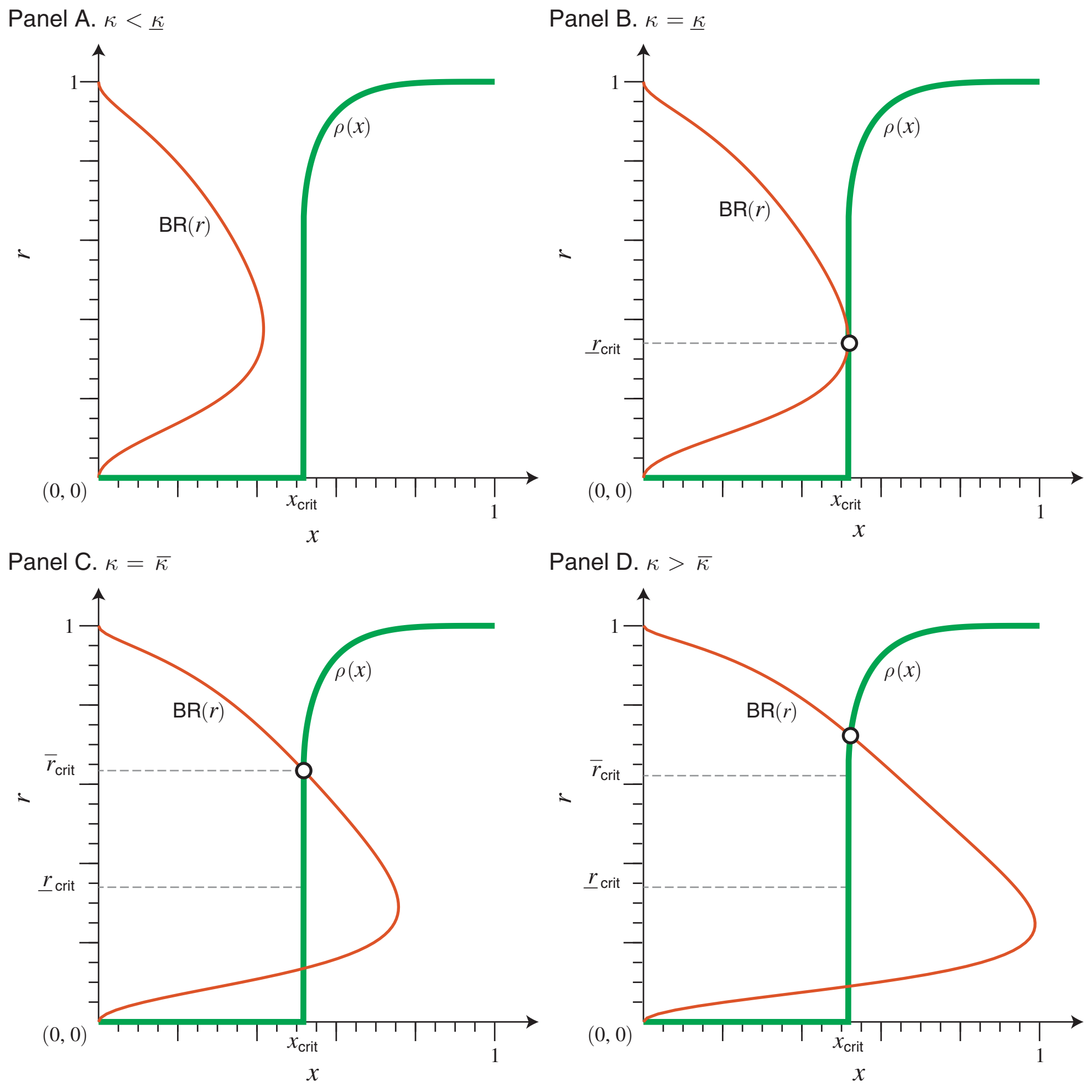

(*Continued*)

FIGURE 6

We now pause to comment on the techniques used to establish these results. While the intuition looks straightforward in our sketches, proving Theorem 1 is technically challenging. We have to get a handle on the shape of the best-response correspondence, and show that the highest intersection of it with the reliability curve moves as depicted in Figure 6. Investment incentives (which determine the shape of the best-response curve, and the implications of changing $\kappa$) are complex, and so equilibrium investment is difficult to characterize directly. We do not have an explicit expression for the best-response curve. It also does not have the kinds of global monotonicity properties that might permit standard comparative statics approaches: investments in relationship strength are strategic complements in some regions of the parameter space, and strategic substitutes in others.[25] However, some control on

[25] When a firm's suppliers are very unreliable, there is little incentive to invest in stronger relationships with them—there is no point in having a working supply relationship when the suppliers cannot produce their goods. On the other hand, when a firm's suppliers are extremely reliable, a firm can free ride on this reliability and invest relatively

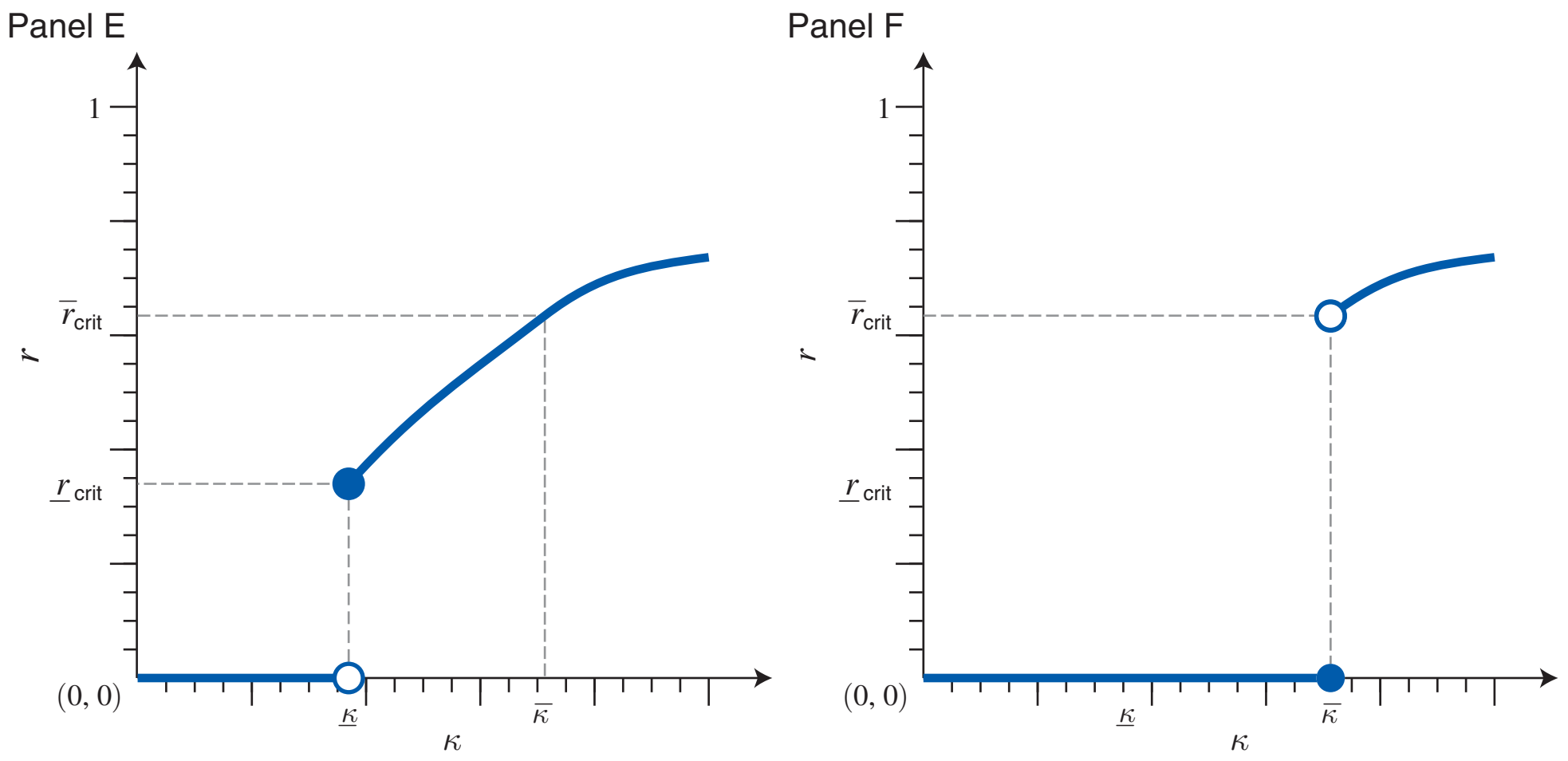

FIGURE 6 (*Continued*)

*Notes:* Panels A–D show, for several different values of $\kappa$, the reliability curve $\rho(x)$, and the best-response curve, a function of the vertical axis variable $r$ (the reliability of a typical supplier). The symmetric undominated equilibrium is marked with a small circle, when a nonzero one is present. Panel A shows a situation with $\kappa < \underline{\kappa}$, where the only equilibrium involves no investment. Panel B shows the symmetric undominated equilibrium (equilibrium) for $\kappa = \underline{\kappa}$. Panel C shows the equilibrium for $\kappa = \bar{\kappa}$. Panel D shows the equilibrium for $\kappa > \bar{\kappa}$. Panel E plots how equilibrium reliability varies with $\kappa$. Panel F shows how reliability varies with $\kappa$ following an arbitrarily small negative shock to institutional quality $\underline{x}$; note the drop on the interval $\kappa \in [\underline{\kappa}, \bar{\kappa}]$.

the slopes of the intersecting curves is necessary to analyze the dependence of the intersection on parameters. A key step in our analysis is showing that if the highest intersection of best response and reliability curves occurs at $r \geq \bar{r}_{crit}$, then the best response curve must be negatively sloped in $r$ there, and strictly so for an intersection at $r > \bar{r}_{crit}$. We do so by establishing certain properties of the system of polynomial equations that (asymptotically) characterizes the intersection. Another crucial step is to show that there is at most one point of intersection above $\bar{r}_{crit}$ in our diagrams. This permits us to focus on a unique outcome of interest, and to sign the effects of moving $\kappa$ unambiguously, which is important for the results on equilibrium fragility.

At a technical level, there are two parts to many of our proofs. One part focuses on an idealized "limit supply network," which, in a suitable sense, has infinite depth and therefore a lot of symmetry (a firm's suppliers look exactly like the firm itself). This symmetric network gives rise to fixed-point expressions that we can manipulate to establish the key facts mentioned above. There is then a separate task of transforming these limit results into statements about the supply network with large but finite depths.

*Comparative Statics in Baseline Relationship Strength.*—We close this subsection by noting a corollary: the comparative statics of equilibrium as the baseline quality of institutions $\underline{x}$ changes are analogous to those documented with respect to $\kappa$ in Theorem 1. Here we explicitly include $\underline{x}$ as an argument in $x^*$.

little in its relationships, knowing that as long as it has one working relationship for each input it requires, it is very likely to be able to source that input.

COROLLARY 1: *Suppose* $\kappa' > \kappa$. *Then, for large enough* $\tau$, *if* $x^*_\tau(\kappa', \underline{x}) > x^*_\tau(\kappa, \underline{x})$, *there is an* $\underline{x}' > \underline{x}$ *such that* $x^*_\tau(\kappa', \underline{x}) = x^*_\tau(\kappa, \underline{x}')$.

## B. *Fragility*

Critical equilibria are important because, as the example in Figure 6 shows, they create the possibility of fragility: small unanticipated shocks to relationship strengths via a reduction in $\underline{x}$ can result in a collapse of production. We formalize this idea by explicitly examining how the supply network responds to a shock to the baseline quality of institutions $\underline{x}$, which for simplicity is taken to have zero probability. (The analysis is robust to anticipated shocks that happen with positive probability; see Section VC.)

DEFINITION 2 (Equilibrium Fragility):

- *There is* ***equilibrium fragility*** *at* $\kappa$ *if for any* $\eta, \epsilon > 0$, *for large enough* $\tau$, *we have*

$$\rho\left(x^*_\tau(\kappa) - \epsilon, \mu_\tau\right) < \eta.$$

  *That is, a shock that reduces relationship strengths arbitrarily little* $(\epsilon)$ *from their equilibrium levels leads to a reliability very close to* 0 *(within* $\eta$*).*
- *We say there is* ***equilibrium robustness*** *at* $\kappa$ *if there is not equilibrium fragility.*

Note that the definition of equilibrium fragility makes sense only if $x^*_\tau(\kappa) > 0$ for all large enough $\tau$, and this implies reliability bounded away from zero before the shock, since otherwise there would be no incentive for investment. In the definition of fragility, while shocking $\underline{x}$, we hold investment decisions and entry choices fixed. This corresponds to the assumption, discussed in Section IVC below, that investments in supply relationships and entry decisions are made over a sufficiently long time frame that firms cannot change the quality of their supply relationships or their entry decisions in response to a shock.

PROPOSITION 3: *Under the conditions of Theorem 1, there is equilibrium fragility at any* $\kappa \in [\underline{\kappa}, \bar{\kappa}]$ *and equilibrium robustness at any* $\kappa > \bar{\kappa}$.

Proposition 3 follows immediately from the definition of equilibrium fragility and our previous characterization.

## C. *Some Comments on Interpretation*

*Sources of Shocks*.—There are many ways in which the small systemic shock involved in our notion of fragility might arise in practice. In the introduction, we discussed the example of a small shock to credit markets making many supply relationships slightly more likely to fail. Another example comes from considering new trade regulations: if there is uncertainty about which relationships will be affected by new compliance issues, each supply relationship will be ex ante more prone to disruption. For a third example, consider an increasing backlog in commercial courts—a circumstance that makes contracts more costly to enforce. This again decreases the probability that contracts function in some states of the world—an uncertainty that can affect many players in the supply chain. A final example comes from supply chain stresses due to macroeconomic shocks. Technological factors, such as social distancing measures imposed during a pandemic, can cause stochastic delays and disruptions throughout the dependency links in a supply network. Congestion provides another type of stress. For example, a demand shock to some goods can require ports and transportation hubs to accommodate more traffic, causing queues. These queues delay shipments of goods in many supply networks, even those not affected by the demand shocks that caused the queues (McLain, Matthews, and Paris 2021).

*Time Scales*.—In interpreting our model, a crucial issue is that of timing: the time scales on which reliability is degraded, and on which decisions are made. We define the short run to be period of time over which firms are *not* able to adjust their relationship strengths. It is this time scale on which the shock we study has its most direct implications in terms of freezing supply chains. Evidence from Barrot and Sauvagnat (2016) suggests that in practice this time scale is on the order of magnitude of several quarters. In contrast, we define the medium run as a period of time over which adjustments in relationship strength are possible.[26] Thus, following a shock to relationship strengths, in the short run a collapse in production will occur when the supply network is in the fragile regime but in the medium run, it is possible that firms can choose investment levels that allow the supply network to resume being productive. Our equilibrium selection here is optimistic: it focuses on the outcome where, in recovering from the shock, firms coordinate on the most productive investment equilibrium, and thus limit the losses. Alternatively, once production has become disrupted, it could be that firms stop investing altogether. If such coordination problems occur, disruptions can be longer-lasting and consistent with long-lasting productivity damage following a shock.

## V. Discussion of the Model

This section discusses our modeling choices. First, in Section VA, we highlight which features of our model are essential and motivate them. We provide some benchmarks to help highlight their importance in Section VB. Beyond the key assumptions, we make a variety of assumptions for tractability or simplicity. In Sections VC and VD, we discuss the robustness of our model to various relaxations of these nonessential assumptions.

### A. *Essential Features*

The three features that are essential in our model are that supply relationships are specific (firm-to-firm) and subject to disruption, that firms endogenously invest to strengthen their relationships, and that production is complex in that it relies on multiple essential inputs at each of a sufficiently large number of layers.

[26] Other technological dimensions, such as the complexity of production, $m$, might change on a longer time scale still; see Section VIID.

There is considerable evidence for the importance of the specific sourcing relationships that are at the heart of the model. These have been documented in many settings: for relationship lending between banks and firms, see Petersen and Rajan (1994, 1995); for traders in Madagascar, see Fafchamps and Minten (1999); for the New York apparel market, see Uzzi (1997); for food supply chains, see Murdoch, Marsden, and Banks (2000); for the diamond industry, see Bernstein (1992); for Japanese electronics manufacturers, see Nishiguchi (1994); and so on. Indeed, even in fish markets, a setting where we might expect relationships to play a minor role, they seem to be important (Kirman and Vriend 2000; Graddy 2006). Specific sourcing relationships in supply networks are also a major concern of the management literature on supply chains (Datta 2017). Barrot and Sauvagnat (2016) find that firms have difficulty switching suppliers even when they need to do so. These literatures examine many reasons behind firms' reliance on a small number of suppliers to source a given type of input. For example, technological compatibility and geography can limit the pool of potential suppliers; hold-up problems can make trust important; and frequent repeated interactions can help firms to avoid misunderstandings.

The specific relationships that firms maintain for sourcing facilitate the contagion of disruption. Indeed, cascading disruptions (see, for example, Carvalho et al. 2021) are evidence of firms' reliance on failure-prone sources of supply that cannot be quickly replaced. Interesting qualitative descriptions of cascades of disruption due to idiosyncratic shocks can be found in the business literature. A fire at a Philips Semiconductors plant in March 2000 halted production, preventing Ericsson from sourcing critical inputs, causing its production to also stop (*Economist* 2006). Ericsson is estimated to have lost hundreds of millions of dollars in sales as a result, and it subsequently exited the mobile phone business (Norrman and Jansson 2004). In another example, two strikes at General Motors (GM) parts plants in 1998 led 100 other parts plants, and then 26 assembly plants, to shut down, reducing GM's earnings by an estimated $2.83 billion (Snyder et al. 2016). Though these particular examples are especially well-documented, disruptions are more frequent than might be expected. In a survey of studies on this subject in operations and management, Snyder et al. (2016, p. 89) write, "It is tempting to think of supply chain disruptions as rare events. However, although a given type of disruption (earthquake, fire, strike) may occur very infrequently, the large number of possible disruption causes, coupled with the vast scale of modern supply chains, makes the likelihood that some disruption will strike a given supply chain in a given year quite high." An industry study recorded 1,069 supply chain disruption events globally during a six-month period in 2018 (*Supply Chain Quarterly* 2018). Elliott and Golub (forthcoming) survey a range of evidence on the importance of specific sourcing in complex supply networks and the impact of disruptions.

Endogenous investment in relationships is the second key model feature. Given the frequency of disruptions and the impact these can have on firms' profitability,[27] it is natural that firms take actions to mitigate them. In practice, these investments are often "soft" in nature. A literature in sociology helps document them—a prominent contribution being Uzzi (1997), which offers a detailed account of the systematic efforts and investments made by New York garment manufacturers and their suppliers to maintain good relations. These investments include practices such as building a better understanding of a supplier's or customer's capabilities by visiting their facilities, querying odd instructions to help catch mistakes, building social relationships that span the organizations, and reciprocal gift giving. Such investments are hard to observe and even harder to verify. They are thus very difficult to contract on. This motivates the way we model investments—as a noncooperative game of private investment in one's local relationships.[28]

[27] Hendricks and Singhal (2003, 2005a, b) examine hundreds of supply chain problems reported in the business press. Even minor disruptions are associated with significant and long-lasting declines in sales growth and stock returns.

The third feature crucial for our theoretical results is that production is complex. One aspect of complexity is in the horizontal dimension of network diagrams such as Figure 1: firms need to procure multiple inputs via specific sourcing relationships. Barrot and Sauvagnat (2016) provide evidence that supports this assumption. They show that if a supplier is hit by a natural disaster, it severely disrupts the production of their customers *and* also negatively impacts their customers' other suppliers. This would be unlikely to occur if production did not rely on multiple specifically-sourced, complementary inputs.[29] Relatedly, this evidence suggests that inventories are not held at high enough levels to avoid disruption, though they may help delay it. Another aspect of complexity is that production takes place in multiple layers—i.e., is sufficiently deep. As we show in Section SA7 of the online Appendix, even without taking any limits, realistic degrees of complexity (both in the horizontal or vertical dimension) generate patterns that have the key qualitative features driving our main results.

A concrete example of a firm with a relatively complex supply chain is Toyota. After the Great East Japan Earthquake, Toyota suffered considerable disruption to its supply network due, in part, to the impact on firms several layers upstream of it. Following this, Toyota made an effort to map out its supply network so it could better anticipate and respond to such problems. It has been reported that this mapping got as far as the tenth level and identified 400,000 items that Toyota sources directly or indirectly (McLain 2021); nevertheless, the supply network was sufficiently vast that the mapping project was not finished.

One final observation is that the nodes in our model can be reinterpreted as describing processes or inputs, rather than firms, as long as the key conditions outlined in this section hold. From this perspective, the phenomena we discuss may happen within a firm or other organization. Increasing the strength of relationships would then correspond to increasing various forms of redundancy or robustness of within-firm sourcing. When agency frictions are significant and decisions about robustness are made in a decentralized way, our model of endogenous investments would also be relevant.

[28] The key feature here is the noncooperative part. Even if firms were allowed to invest in others' relationships, rather than merely their own, they would typically not want to in equilibrium. Following the logic of Bergstrom, Blume, and Varian (1986), each firm would be willing to invest in each relationship (between any two firms) up to the point that the marginal private benefit of the investment equals the marginal cost. However, the marginal benefits are heterogeneous and it will typically be possible for this condition to be satisfied by only one firm—the one that experiences the highest returns from such investment.

[29] Indeed, other suppliers providing substitute inputs would benefit from the disruption to a competitor rather than being adversely affected.

### B. *Contrasts with Models without the Essential Features*

In this subsection, we briefly examine two alternative models as benchmarks that help highlight the necessity of key features discussed in the previous section.

*Contrast I: Sourcing through Spot Markets*.—In this benchmark, each firm sources all its inputs through spot markets, rather than requiring pre-established relationships. A spot market is populated by those potential suppliers that are able to successfully produce a certain input. To keep the technology of sourcing comparable to that of the main model, we posit that, after a buyer places an order in a spot market, there is still a chance that sourcing fails. (A shipment might be lost or defective, or a misunderstanding could lead the wrong part to be supplied.) In other words, we now assume each firm extends relationships *only* to functional suppliers (as opposed to suppliers whose functionality is random, in our main model), but we still keep the randomness in whether the sourcing relationships work. Two other features are carried over from the main model: firms may multisource, extending multiple such relationships; and the randomness in sourcing is realized simultaneously, independently across relationships.

We now work out which firms can produce, focusing on an example where each firm requires two inputs ($m = 2$) and each firm *if* multisources by contracting with two potential suppliers of each input ($n = 2$) in the spot market. Let the probability that a given attempt at sourcing an input succeeds be $x$. The probability that both potential suppliers of a given input type fail to provide the required input is $(1-x)^2$, and the probability that at least one succeeds is $1-(1-x)^2$. As the firm needs access to all its required inputs to be able to produce, and it requires two different input types, the probability the firm is able to produce is $\left(1-(1-x)^2\right)^2$. In Figure 7, we plot the analogue of reliability—the probability a random firm is able to produce—as a function of the success probability of individual sourcing attempts. Reliability increases smoothly as $x$ increases. This benchmark thus shows that perfect spot markets remove the discontinuities in our main analysis.

On the other hand, the fragility of complex production remains a concern in the presence of *imperfect* substitutes for specifically sourced inputs. More precisely, suppose the quality of a firm's output is discounted by a certain factor whenever the firm cannot source high-quality inputs for all customized inputs required. Such a firm instead produces a lower-quality version of its good. Shocks disrupting sourcing relationships propagate downstream just as in our main analysis, but affected firms can now produce low-quality goods.

*Contrast II: Sourcing for Simple Production*.—To emphasize that it is essential that multiple inputs are sourced through relationships, we consider a benchmark model where each firm requires only a single relationship-sourced input ($m = 1$). Because each firm requires only *one* type of risky input relationship to work, we call such production *simple*.[30] We plot how the probability of successful production

[30] As a matter of interpretation, there may be more than one physical input at each stage. The key assumption is that all but one are sourced as commodities rather than through relationships, and so are not subject to disruption via shocks to these relationships. The term "simple" is inspired by the contrast between simple and complex contagion

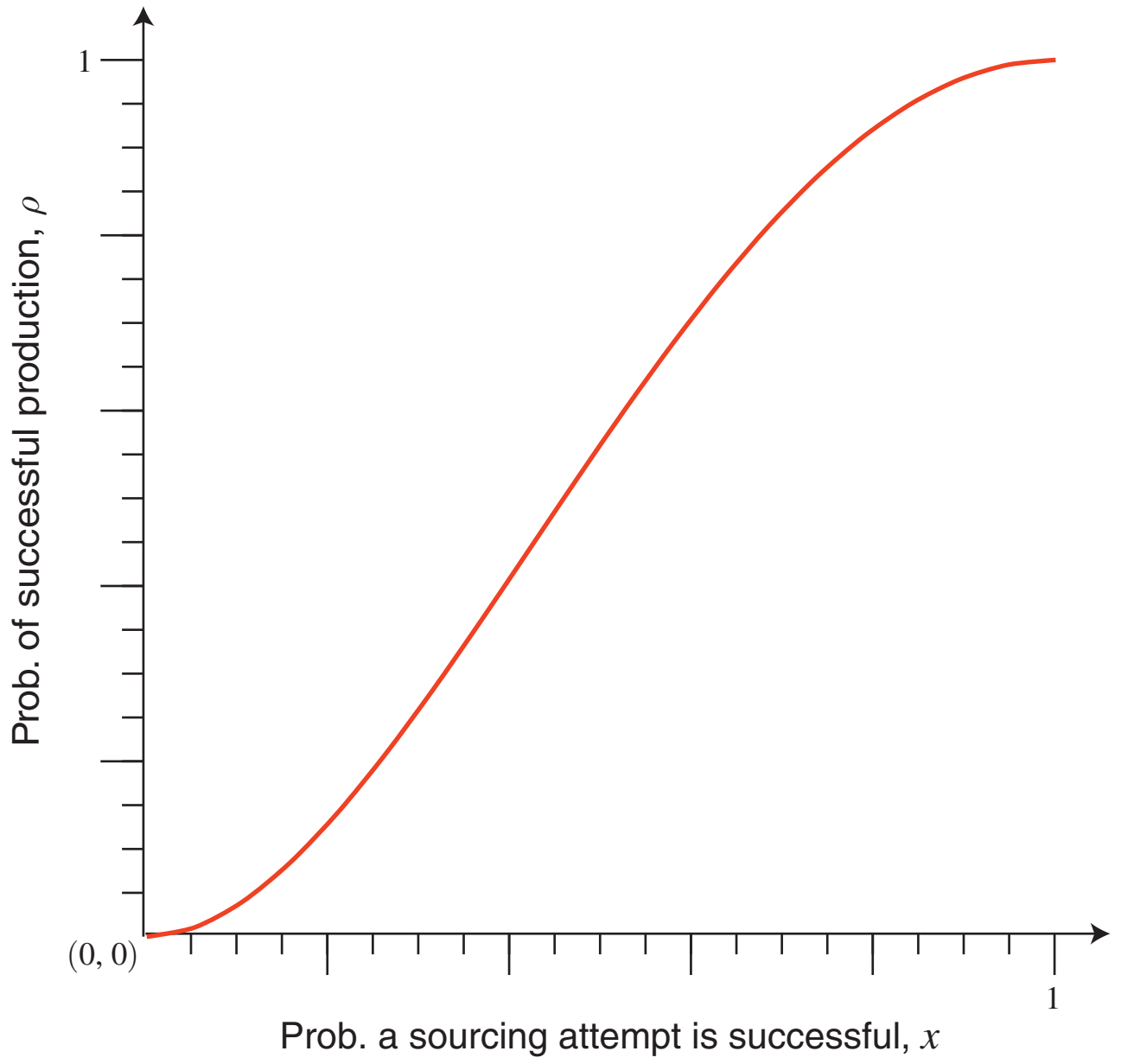

FIGURE 7

*Note:* The probability of successful production for a firm as market-based sourcing attempts become more likely to succeed.

varies with relationship strength in Figure 8 panel B for an example with $n = 2$. This may be compared to the case of complex production illustrated in Figure 4 panel B. For values of $x < 0.5$ the probability of successful production is 0, and for values of $x > 0.5$ the probability of successful production is strictly positive, paralleling the qualitative transition seen in the main model. There is a stark contrast, however, in the nature of the change at the value of x for which the probability of successful production first becomes positive; productivity is continuous in the simple-production case, with no sharp jump (though the derivative is discontinuous).

One can obtain the reliability curve via a graphical analysis analogous to that of Section IIC, now with the $\mathcal{R}_x(r)$ curve depicted in Figure 8 panel A replacing the curve shown in Figure 4 panel A. The key difference is that, because the curve $\mathcal{R}_x(r)$ is globally concave, as $x$ decreases and the curve descends, the points of intersection depicted in Figure 8 panel A approach $(0,0)$ continuously in $x$. This corresponds to the continuity of the curve in Figure 8 panel B.

Another intuition for the transition in reliability in this case is familiar from the networks literature and in particular from studies of contagion; see, for example, Elliott, Golub, and Jackson (2014) in the context of financial contagion, or Sadler (2020) for information diffusion. In the large-depth limit, production will be successful if the supply tree of functional producers upstream of a given firm continues indefinitely, rather than being extinguished due to an excess of failures. This depends on whether the rate at which new branches in the network are created is higher or in the literature on networks: see Elliott and Golub (forthcoming) for a discussion of this in the context of supply network models.

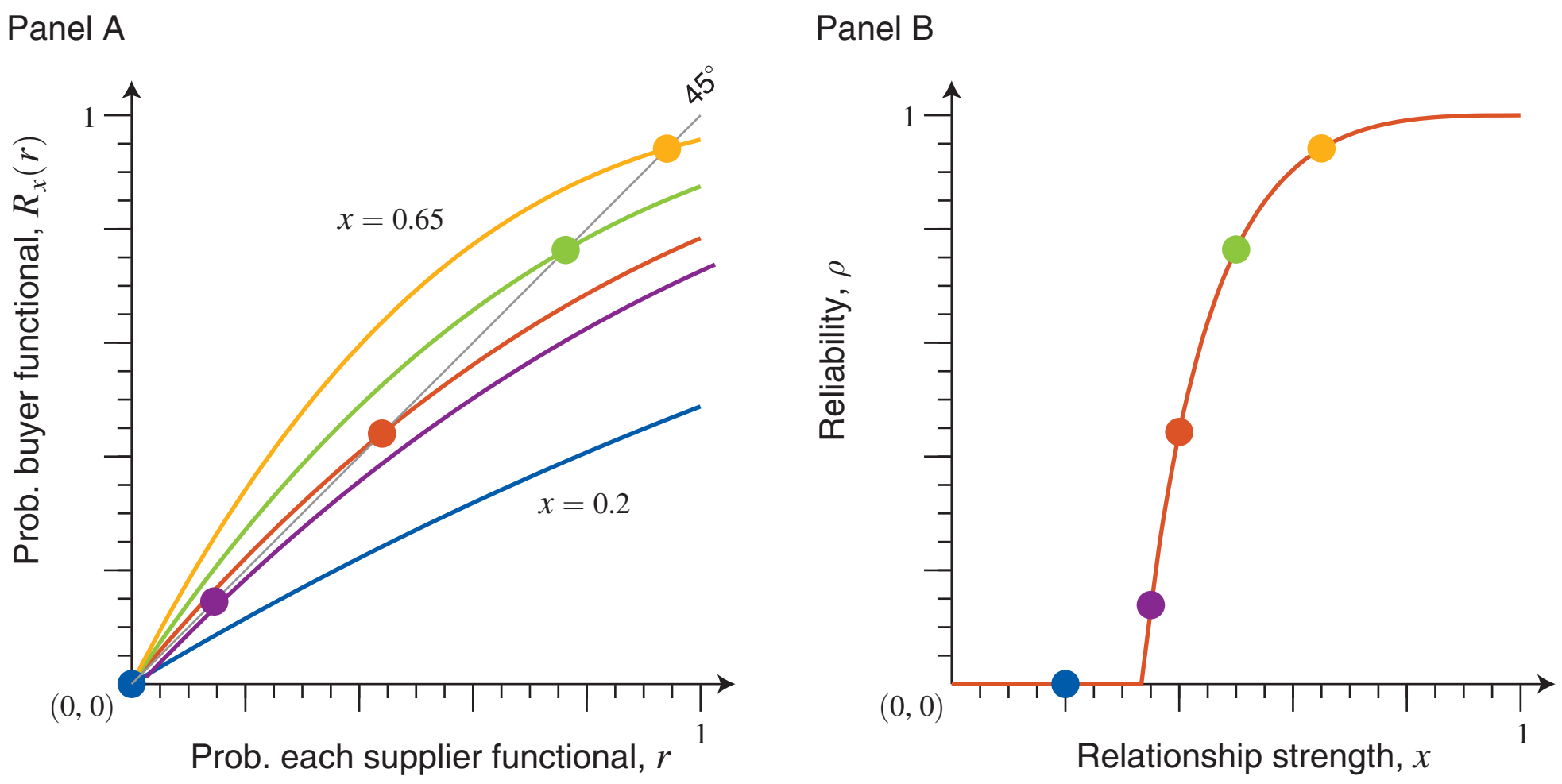

Figure 8

*Notes:* Panel A: the probability, $\mathcal{R}_x(r)$, that a focal firm is functional as a function of $r$, the probability that a random supplier is functional. This plot is for $n = 3$ with $x \in \{0.2, 0.35, 0.4, 0.5, 0.65\}$. The plot parallels Figure 4 panel A, which depicts the same function for the complex production $(m > 1)$ case. The marked intersections with the 45 degree line reflect limit reliability outcomes. Panel B: the reliability in a supply network for simple goods $(m = 1$ distinct inputs needed) as relationship strength $x$ varies.

lower than the rate at which existing branches die out due to failure. It turns out that when $x > 0.5$, a supply tree grows without bound in expectation, while when $x < 0.5$ it dies out.[31] The kink in the probability of successful production around the threshold of 0.5 is related to the emergence of a giant component in an Erdös-Rényi random graph. That continuous phase transition is different from the discontinuous one driving our main results. To see the importance of the difference in these phase transitions for equilibrium fragility, consider a productive equilibrium when production is simple. It can be immediately be seen from Figure 8 panel B that a small shock to reliability $x$ will cause only a small change in the probability of successful production. The reason is that, in contrast to the complex production case, the reliability curve does not become arbitrarily steep in the large-depth limit.

### C. *Robustness: Extensions Relaxing Simplifying Assumptions*

In this section we discuss directions in which the analysis can be extended to show that certain simplifying assumptions are not essential to the main findings. We begin with extensions to the model in which firms remain homogeneous (in the distribution of their network positions), and then turn to heterogeneities in the next subsection.

[31] Given that each producer has two potential suppliers for the input, and each of these branches is operational with probability $x$, the expected number of successful relationships a given firm in this supply network has is $2x$. When $x < 1/2$, each firm has operational links to less than 1 supplier on average, and so the rate at which branches in the supply tree fail is faster than the rate at which new branches are created. The probability that a path in the supply network reaches beyond a given length $l$ then goes to 0 as $l$ gets large and production fails with probability 1. On the other hand, when $x > 1/2$, the average number of operational links each firm has exceeds 1, and so new branches appear in the supply tree at a faster rate than they die out, leading production to be successful with strictly positive probability.

*Investments on the Extensive Margin*.—Our modeling of investments in supply relationships is compatible with multiple interpretations. The first interpretation is that the set of possible suppliers is fixed, and the investment works on the intensive margin to improve the quality of these relationships (e.g., by reducing misunderstandings and so on). The second interpretation is that the investment works on the extensive margin; that is, firms work to find a supplier willing and able to supply a given required input type, but their success in this search is stochastic. In this interpretation there is a fixed set of $n$ potential suppliers capable of supplying the required input, and each one of them is found independently with probability $x_{if}$.[32] Conditional on a supplier being found, the relationship is operational. In online Appendix SA9 we discuss a richer extensive-margin interpretation, as well as an extension that permits separate efforts to be directed to the extensive and intensive margins simultaneously.

*Anticipated Shocks*.—We assume that the small shock to $x$ is unanticipated. It is straightforward to use our analysis to see that this is not essential. Suppose now all firms anticipate that a shock to $x$ will happen with some probability $p_{\text{shock}}$, and suppose $\kappa$ is in an open subset of the critical productivities, so that a collapse would have occurred in the baseline model. When we write the profit function for firms in the extended model, there will now be an expectation over the shock's arrival which will change profits and the best-response correspondence. However, for a small enough probability of the shock, this change will be small, and the best-response curve in a plot such as Figure 6 panel C will move only slightly. Thus the equilibrium will still be on the precipice if it was there before. Indeed, for $\kappa$ not too close to the boundary of the critical productivities, the probability of the shock can be quite substantial without changing the main prediction. The key intuition is as follows. Because of externalities, all firms may prefer a commitment to invest more to avoid being on the precipice (and avoid disruptions due to the shock). But the free-riding problem can be too severe to achieve this, even when shocks are anticipated. Indeed, note that in Figure 6, even fairly large shifts in the best-response curve can leave the equilibrium on the precipice, implying that these free-riding effects are considerable.

*Shocks to Firms*.—The fundamental source of shocks in our model is at the level of links in the supply network. One could also consider shocks directly hitting firms—i.e., shocks to nodes in our supply network rather than links. The key conclusions regarding precipices are robust to adding this source of shocks, or even making node-level shocks the main source of disruptions.[33] We focus on shocks to relationships for simplicity, but the mathematical forces we identify also operate in alternative models of disruption.

[32] This search technology is similar to ball-and-urn models of search, and is compatible with a matching function exhibiting constant returns to scale (see, for example, Hall 1979).

[33] This variant of the model is worked out in Elliott and Golub (forthcoming).

*Endogenous Entry*.—Our model takes the varieties in the market as given, and does not include entry decisions by firms. This is done to keep the model simple. However, the key insights carry over to a model with entry.

Consider an extended model where there is an entry stage preceding all others. At this stage, firms $(i,f)$ simultaneously decide whether to pay a sunk entry cost $\Phi(f)$, where $\Phi$ is a strictly increasing function. Then the measure of varieties in each product is set equal to the measure of entering firms (i.e., each entering firm produces one variety), and the "investment game" where firms choose their relationship strengths proceeds as we described earlier. A (nontrivial) symmetric equilibrium is now one where $\bar{f} > 0$ firms enter in each product and the expected postentry profit of the marginal firm is equal to its entry cost.

This model also features a nonempty open set of productivity parameters $\kappa$ where equilibrium production is on the precipice. The basic logic is as follows. When the supply network is reliable and gross profits after entry are high, firms want to enter. As they enter, competition drives down gross profits and makes it less appealing to pay costs to make relationships strong. Recalling Figure 6, the key to the argument is that increasing entry moves the best-response curve leftward for a given $\kappa$. So, in equilibrium, relationships get weaker as more entry occurs. The question is where this dynamic stops. The precipice is a natural stopping point. Once the investment level is $x_{\text{crit}}$, reliability can adjust down until further entry is deterred, while investment remains essentially the same. By reasoning parallel to that of our main results, a fragile state is reached for an open set of model parameters.

## D. *Robustness to Heterogeneity*

We have so far assumed for tractability that all firms have identical network positions ex ante. One might suspect that the regularity of the network structure, or some other kind of homogeneity, plays an important role in generating the sharp transition in the probability of successful production. However, homogeneities across the firms are not essential to our main points. To establish this, we now discuss extensions with two forms of heterogeneity in firm type.

*Partial Knowledge of Depth*.—We have assumed that firms make investments ex ante of the realization of their supply network depths, so that their investment decision problems are identical. This symmetry simplifies the analysis considerably. However, in practice, firms do have some information about the depth of their supply networks. We now introduce this into the model and give conditions under which our main findings extend.

We refer to the model analyzed in Sections II–IV as the *baseline model*, and we will define the model with partial knowledge of depths with reference to it.

The firms are now partitioned into two types: low-depth firms, with depth less than or equal to $\bar{d}_l \geq 0$, constitute a proportion $\mu_l < 1$ of the population, and high-depth firms, with depth greater than $\bar{d}_l$, constitute the remainder. (We discuss the case of more than two types below.) Firms' information about depth consists of knowing which type they are. We use the subscript $l$ for low-depth firms and $h$ for the high-depth firms.

Each type of firm has gross profits from production equal to $g_t(r_t)$, where $r_t$ is the aggregate reliability of that type of firm ($t = l$ for low-depth firms, and $t = h$ for high-depth ones). This amounts to assuming that each type of firm sells into a separate market, and its gross profits are dependent on competition in that market through $r_t$. Similarly, each type of firm has a cost function $\frac{1}{\kappa_t} c_t$. The model primitives are otherwise the same, with the functions $g_t$ and $c_t$ satisfying the assumptions made on $g$ and $c$, respectively, in the homogeneous model.

We now extend the definition of equilibrium to allow low-depth and high-depth firms to behave differently; equilibrium again refers to a symmetric Nash equilibrium that is not Pareto-dominated by any other symmetric equilibrium, but "symmetric" now means that strategy depends only on type. We let the vector $\mathbf{x}^*_\tau(\kappa_l, \kappa_h)$ denote relationship strengths for the low-depth and high-depth firms in such an equilibrium (where $\tau$ is, as before, the index on the depth distribution $\mu_\tau$). The function $\rho_t(\mathbf{x}, \mu)$ gives the reliability of type-$t$ firms under strengths $\mathbf{x}$ and depth distribution $\mu$.

Our focus will be on the high-depth firms. We thus define an equilibrium to be fragile if, for large enough typical depth, a very small shock causes the *high-depth firms* to have very low reliability.

DEFINITION 3 (Equilibrium Fragility with Partial Knowledge of Depth): *There is* ***equilibrium fragility*** *at* $\boldsymbol{\kappa} = (\kappa_l, \kappa_h)$ *if for any* $\eta, \epsilon > 0$, *for large enough* $\tau$, *we have*

$$\rho_h\big(\mathbf{x}^*_\tau(\boldsymbol{\kappa}) - \epsilon\mathbf{1}, \mu_\tau\big) < \eta.$$

That is, a shock that reduces relationship strengths arbitrarily little ($\epsilon$) from their equilibrium levels leads to a reliability very close to 0 (within $\eta$) for the high-depth firms.

PROPOSITION 4: *Suppose there is equilibrium fragility at* $\kappa$ *in the baseline model under gross profit function* $g$. *Then in a partial knowledge of depth model with* $g_h = g$, *the following holds: if* $x^*_h(\kappa_l, \kappa) > 0$ *for all sufficiently large* $\tau$ (*i.e., there is positive investment*), *then there is also equilibrium fragility.*

To gain some intuition for the result, note that in a (symmetric undominated) equilibrium, the high-depth firms take the reliability of the low-depth firms as given. In particular, all low-depth firms from which the high-depth firms source are at depth exactly $\bar{d}_l$ and hence have the same reliability. For $\kappa_h = \kappa$, the high-depth firms are then in an equivalent position to firms in our baseline model, except that the firms at depth $\bar{d}_l$ play the role of the depth-0 firms in the baseline model. In that model, the depth-0 firms have reliability 1. In the partial depth knowledge model, the depth $\bar{d}_l$ firms have some reliability $\bar{r} \leq 1$. Nevertheless, as long as $\bar{r} \geq \bar{r}_{\text{crit}}$, this difference does not matter for $\tau$ sufficiently high. This is because the high-depth firms will have the same limit reliability as in the baseline model if they make the same investments, by the reasoning in Section IIC. Our assumption that there is equilibrium fragility at $\kappa$ in the baseline model implies that, in that model, the limit of equilibrium investment is $x_{\text{crit}}$. The proof shows that there will also be limit investment $x_{\text{crit}}$ for the high-depth firms in the symmetric undominated equilibrium of the extended model (as $\tau \to \infty$).

This leaves the case when the reliability of the depth $\bar{d}_\ell$ firms is $\bar{r} < \bar{r}_{\text{crit}}$. We just need to show that in this case limit reliability is either 0 or $\bar{r}_{\text{crit}}$. Hence, it is sufficient to rule out limit equilibrium investment $x > x_{\text{crit}}$. Such investment cannot be supported in equilibrium for the same reasons that it could not be supported in the corresponding baseline model.

It is important for interpreting Proposition 4 that we know the high-depth firms sometimes make positive investments. To show that the positive-investment assumption can be satisfied, note that we can always set $\kappa_l$ sufficiently high to induce equilibrium investments for the low-depth firms above $x_{\text{crit}}$, which is sufficient for ensuring the reliability of depth $\bar{d}_l$ firms is above $r_{\text{crit}}$. Then the conditions on $\kappa_h$ for positive investment parallel those of the baseline model.

When there is equilibrium fragility, it is only the high-depth firms that are affected. Moreover, the low-depth firms operate independently of the high-depth firms, and so even if production is disrupted for all the high-depth firms, the low-depth firms can still produce. This readily extends to a finer partitioning of firms by their depth. In this case, those firms in all but the highest-depth group have bounded depth and behave like the low-depth firms; those in the highest-depth group behave like the high-depth firms. Extending the model to include partial knowledge of depth, as we have done in this section, changes the quantitative implications—fewer firms in a supply chain will be on the precipice—but not the qualitative implications insofar as a positive mass of the firms will be on the precipice. The discussion in Section VA suggests that a considerable number of firms producing high-value products would be included in this set.

It is worth noting that, for simplicity, we have firms depend on others upstream of themselves (their suppliers, their suppliers' suppliers, and so on) but not downstream.[34] If the profits of low-depth firms were instead derived from selling to high-depth firms, then the inability of the high-depth firms to produce would, via reduced demand for the low-depth firms' products, affect the profitability of production for the low-depth firms and they might not be able to profitably produce, either. This type of propagation could be incorporated into the specification of the gross profit function $g$. Baqaee (2018) offers a detailed study of such contagion of failure in both directions in a different model.

*Heterogeneous Product Networks*.—We now consider an extension of our model with flexible heterogeneity in network structure. We focus in this subsection on the mechanics of how reliability depends on relationship strengths. In online Appendix SA8 we flesh out the details rigorously and also present an extension of the endogenous investment model.

The key point in this extension is that the number of inputs required, $m$, and the number of potential suppliers, $n$, are no longer held fixed across the supply network; these are allowed to vary with the product. More formally, there is a finite set of products, $\mathcal{I}$. Each product $i \in \mathcal{I}$ is associated with a *product complexity* $m_i$ and a finite set of inputs $I(i) \subseteq \mathcal{I}$ of cardinality $m_i$. Thus, the number of inputs required can be different for different products. The neighborhoods $I(i)$ define the *product dependence graph*, with the convention that arrows are directed upstream. For each firm producing product $i$, each needed input $j$ can be supplied to the firm by $n_{ij}$ potential suppliers. For each pair $i,j \in \mathcal{I}$, there is a relationship strength $x_{ij}$ such that every link from a firm producing $i$ to a potential supplier of product $j$ has a probability $x_{ij}$ of being operational. The product dependence graph is illustrated in Figure 9 panel A, while the heterogeneities in multisourcing possibilities and in link strength are illustrated in panel B. We denote by $\mathbf{x} = (x_i)_{i \in \mathcal{I}}$ the profile of relationship strength vectors for the different products, where $x_i = \left(x_{i,j}\right)_{j \in I(i)}$.

[34] In our microfoundations (online Appendix SA3), the upstream but not downstream dependence is modeled by letting all firms sell to final consumers, and letting all profits be derived from such sales, while sales of intermediates are at cost.

Analogously to our main model, we introduce depths of each variety. Depth-0 varieties are those that require no specifically sourced inputs. We specify the matching process so that each variety sources from varieties of strictly smaller depth. Subject to this, we allow a flexible distribution of input depths, relaxing the requirement from the baseline model that all suppliers of a given firm are exactly one level less deep. We are again interested in the limit as depths of most firms become arbitrarily large. In online Appendix SA8 we present the details of this setup and characterize this limit, paralleling Section IIB in the homogeneous case. This analysis yields a *limit reliability correspondence* $\boldsymbol{\rho}$ mapping $\mathbf{x}$ to profiles $\left(\rho_i\right)_{i \in \mathcal{I}}$ of reliabilities. The correspondence has the property that, for any $\mathbf{x}$, if depths are sufficiently large, the reliabilities are arbitrarily close to those given by $\boldsymbol{\rho}$.

Our first main result is that the sharp transition in the production function persists in the heterogeneous model. There are values of the vector $\mathbf{x}$ where arbitrarily small shocks to relationship strengths result in large drops in reliability. The following definition is helpful for making this point:

DEFINITION 4 (Critical Product): *The product i is* ***critical*** *at* $\mathbf{x}$ *if the product has positive reliability so that* $\rho_i(\mathbf{x})$ *contains positive values at strengths* $\mathbf{x}$, *but has reliability* $\rho_i(\mathbf{x}') = 0$ *at any profile* $\mathbf{x}'$ *of strengths that is elementwise strictly lower than* $\mathbf{x}$.

In Section SA8.1 of the online Appendix, we generalize Proposition 1 by giving conditions guaranteeing that *critical products exist* in the heterogeneous model. The key condition is, as before, that production requires $m_i \geq 2$ distinct inputs at each step.

We next show that supply networks in the presence of heterogeneities feature a *weakest link* property (Proposition 5). When one product is critical, all products that rely directly or indirectly on it are critical as well.[35] This result uses the notion of a product dependence graph, whose nodes are a set $\mathcal{I}$ of products, and products have directed links to the input products they require (see Figure 9 panel A for an illustration).

[35] Thus, as in Bimpikis, Ehsani, and Ilkılıç (2019), the social planner has different gains from intervening in different parts of the network (see their Proposition 8).

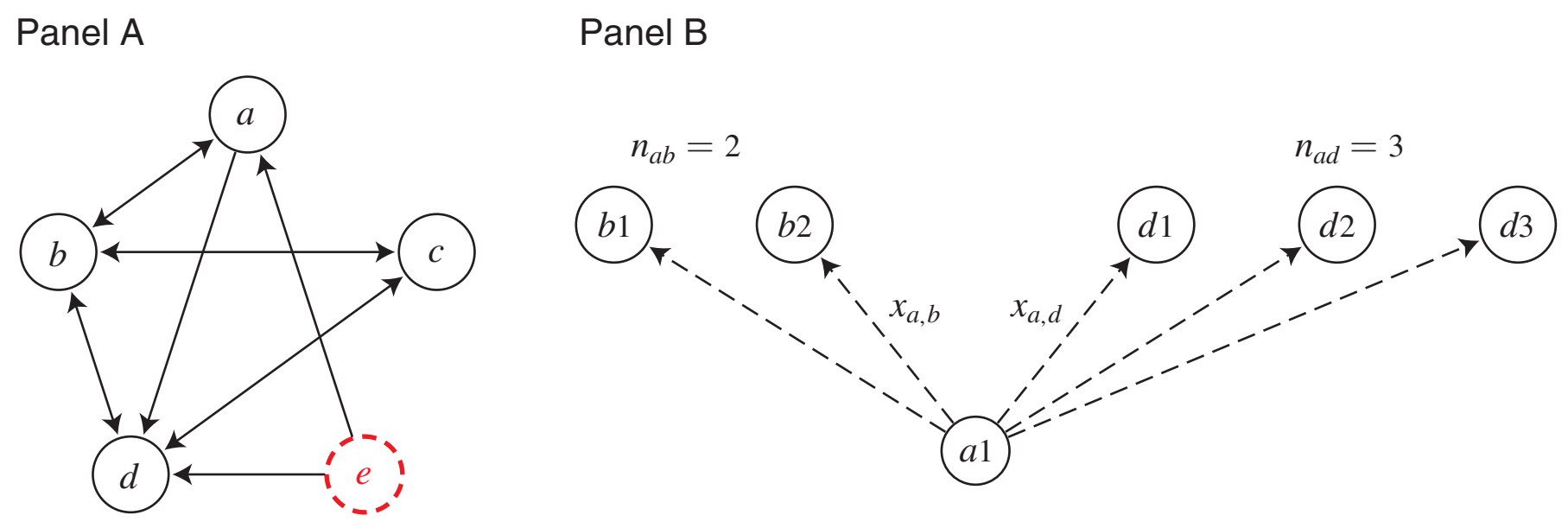

FIGURE 9

*Notes:* Panel A depicts a product dependence graph in the heterogeneous case. An arrow from $a$ to $b$ means that $b$ is an input required in the production of $a$, i.e., $b \in I(a)$. Product $b$ requires $m_b = 3$ inputs (products $a$, $c$ and $d$). Other products require two inputs and thus $m_i = 2$ for $i \in \{a,c,d,e\}$. There are two strongly connected components: $\{a,b,c,d\}$ and $\{e\}$ (which is highlighted). Panel B shows a potential supply network for variety $a1$. Here input products $b$ and $d$ are required, as indicated in panel A, and the number of potential suppliers of each input can be different: there are $n_{ab} = 2$ potential suppliers of input $b$ while there are $n_{ad} = 3$ potential suppliers of input $d$. Moreover, the relationship strength is input-specific, with $x_{a,b}$ being possibly different from $x_{a,d}$.

PROPOSITION 5 (Weakest Link Property):

(*i*) *Suppose product i is critical. Then any product that has a directed path to i in the product dependence graph is critical.*

(*ii*) *Let $\mathcal{I}^{SC} \subset \mathcal{I}$ be a set of products that are part of a strongly connected component of the product dependence graph. Then either all producers of products $i \in \mathcal{I}^{SC}$ are critical or no producers of products $i \in \mathcal{I}^{SC}$ are critical.*

Proposition 5 asserts that supply networks suffer from a weakest link phenomenon. First, in part (i), we see that if a product is critical, then any small systemic shock (which causes it to fail, by definition of criticality) causes the failure of all other products that use it as an input, directly or indirectly. For example, in Figure 9 panel A, we see that if product $b$ is critical, then a small systemic shock to relationship strengths causes its production to fail and will also cause products $a$, $c$, $d$ and $e$ to fail. Products $a$, $c$, $d$ fail because they use $b$ directly as an input. Product $e$ fails because it indirectly uses product $b$ as an input (through $a$ and $d$).

Second, in part (ii), we begin by recalling that a *strongly connected component* of the product dependence graph is a set of nodes (i.e., products) where each node can be reached from any other node in that set by following some directed path of dependencies. For example, in the product dependence graph of Figure 9 panel A, the nodes with solid borders ($a$, $b$, $c$, and $d$) form such a strongly connected component. Node $e$ is not part of this component since it cannot be reached from the other nodes by following the direction of the arrows. A strongly connected component of products is a set of products such that each product is used as input (directly or indirectly) by every other product of that set. This is key to the intuition behind the fact that such a component is only as strong as its weakest link. Indeed, if one product is critical, then all products in that component are also critical at the same time.

In Section SA8.3 of the online Appendix, we provide supporting material for this section and illustrate both the discontinuities and the weakest link property with numerical examples. We also extend our endogenous investment model and show that the configurations we have described are consistent with strategic investment in relationship strength.

## VI. Related Literature

We have already discussed many prior works that are relevant for motivating our assumptions or interpreting our results. In this section we review several high-level connections to related literatures not covered by our previous discussions.

There has been considerable recent interest in markets with nonanonymous trade mediated by relationships.[36] The work most closely related to ours in this area also studies network formation in the presence of shocks. This includes work in the context of production (e.g., Fafchamps 2002; Levine 2012; Brummitt et al. 2017; Bimpikis, Candogan, and Ehsani 2019; Yang, Scoglio, and Gruenbacher 2019; Amelkin and Vohra 2019), work on financial networks (e.g., Cabrales, Gottardi, and Vega-Redondo 2017; Elliott, Hazell, and Georg 2018; Erol 2018; Erol and Vohra 2018; Jackson and Pernoud 2019), and other contexts (e.g., Blume et al. 2011, Jackson, Rodriguez-Barraquer, and Tan 2012; Talamàs and Vohra 2020). More broadly, the aggregate implications of relying on relationships to transact have been studied across a variety of settings.[37] At a methodological level, we advocate an approach that may be useful more broadly: agents make a continuous choice that determines the probability of their relationships operating successfully. The links that form may, however, fail in a "discrete" (i.e., discontinuous) way. The first feature makes the model tractable, while the second one brings discrete properties of networks into the model; this ultimately leads to the possibility of discontinuities in the aggregate production function and distinguishes the predictions from models where the aggregate production function is differentiable. It might be thought that aggregating over many supply chains, these discontinuities would be smoothed out at the level of the macroeconomy; we show they are not.

There is a vibrant literature in macroeconomics on production networks. This literature dates back to investigations of the input-output structure of economies and the implications of this (Leontief 1936). Carvalho and Tahbaz-Salehi (2019) provide a comprehensive survey. Three recent developments in the literature are particularly relevant to our work: (i) the modeling of the endogenous determination of the input-output structure; (ii) a firm-level approach as opposed to considering interindustry linkages at a more aggregated level; (iii) an interest in nonlinearities in aggregate production. Some of the most relevant work on (i) and (ii) includes Atalay et al. (2011); Oberfield (2018); Carvalho and Voigtländer (2014); Acemoglu and Azar (2020); Boehm and Oberfield (2020); Tintelnot et al. (2018); Liu (2019);

[36] A literature in sociology emphasizes the importance of business relationships; see for example Granovetter (1973) and Granovetter (1985). For a survey of related work in economics see Goyal (2017).

[37] For work on thin financial markets see, for example, Rostek and Weretka (2015); for buyer-seller networks see, e.g., Kranton and Minehart (2001); and for intermediation see, e.g., Gale and Kariv (2009). Our work focuses on network formation for *production* and emphasizes the distinctive network formation concerns that arise due to strong complementarities.

Acemoglu and Tahbaz-Salehi (2020); König et al. (2019); Taschereau-Dumouchel (2022); and Kopytov et al. (2021). On the third point, Baqaee and Farhi (2019, 2020) focus specifically on the implications of nonlinearities, and discuss how nonlinearities in firms' production functions propagate and aggregate up. Whereas they focus on smooth nonlinearities, we show that especially extreme nonlinearities—discontinuities—naturally come from complex supply networks. We have discussed throughout how our explicit modeling of discrete sourcing failures in complex production at the micro level gives rise to new effects (see Section VB).

Strong complementarities in production are crucial for creating fragility in our model, and this aspect of our work builds on a large literature following the seminal contribution of Kremer (1993). This literature argues that complementarities can help provide a unified account of many economic phenomena. These include very large cross-country differences in production technology and aggregate productivity; rapid output increases during periods of industrialization; and the structure of production networks and international trade flows; see, among many others, Ciccone (2002); Acemoglu, Antràs, and Helpman (2007); Levchenko (2007); Jones (2011); and Levine (2012).[38] A possible concern with this literature is that if firms were allowed to take actions that mitigate supply risks (for example, by multisourcing), this would endogenously dampen the crucial complementarities. Our work helps address this concern. We show that even when firms can take actions to mitigate their risk by multisourcing, a very severe form of equilibrium fragility arises, and strong complementarities continue to manifest in the aggregate production function.

At a technical level, our work is related to a recent applied mathematics literature on so-called *multilayer networks* and their phase transitions (Buldyrev et al. 2010). Discontinuities such as the one we study are termed first-order phase transitions in this literature.[39] Buldyrev et al. (2010), and subsequent articles in this area such as Tang et al. (2016) and Yang, Scoglio, and Gruenbacher (2019), study quite different network processes, typically with exogenous networks. We show that discontinuous phase transitions arise in canonical models of production networks, once specific sourcing relationships are taken into account. Moreover, in contrast to prior works that took supply networks and shocks to be exogenous, we endogenize investments affecting the probabilities of link operation and elucidate a new economic force endogenously putting equilibria on a precipice. Dasaratha (forthcoming) studies endogenous investment to form a random network in an information-sharing model without aggregate shocks; that work identifies interesting and distinct interactions between phase transition phenomena and equilibrium network formation. Predating the recent literature just discussed, Scheinkman and Woodford (1994) used insights from physics models on self-organized criticality to provide a "sandpile" model of the macroeconomy in which idiosyncratic shocks have large aggregate effects.[40] The setup and behavior of their model are rather different from ours: the main point of commonality is in the concern with endogenous fragility. In our model, investments in supply relationships leave the supply network robust to idiosyncratic shocks, but very sensitive to arbitrarily small systemic shocks to relationship strength.

[38] Prior to his literature, Jovanovic (1987) examines how strategic interdependencies or complementarities can produce aggregate volatility in endogenous variables despite only seemingly "diversifiable" idiosyncratic volatility in exogenous variables.

[39] These can be contrasted with second-order phase transitions, such as the emergence of a giant component in a communication network, which have been more familiar in economics—see Jackson (2008).

[40] Endogenously, inventories reach a state analogous to a sandpile with a critical slope, where any additional shock (grain dropped on the sandpile) has a positive probability of leading to an avalanche.

## VII. Concluding Discussions

We conclude by sketching some implications of our analysis. We first note some basic welfare implications of the model, complementing the descriptive analysis of fragility that has been our focus so far. We then turn to the implications for policy approaches aimed at improving the robustness of supply networks. Finally, we briefly suggest some potential applications of the precipice phenomena to macroeconomic modeling and to studies of industrial development.

### A. *Welfare Implications*

In Section IV we found that decentralized investment choices often result in fragile supply networks. In Section IIIA we showed that a planner would never choose investments that result in the supply network being fragile. These observations imply that, at least when there is equilibrium fragility, the outcome is inefficient. The reason is that a small improvement in reliability from a uniform increase in investments yields arbitrarily large marginal benefits. We now consider the efficiency of equilibrium in our model more broadly.

Our main result in this section is that there is *always* underinvestment in relationship strengths in a productive equilibrium.

COROLLARY 2: *Suppose we have a symmetric equilibrium with relationship strength x. Then*[41] $x < x^{\mathrm{SP}}(\kappa,\mu)$.

This result follows from the proof of Lemma SA4 in the online Appendix, which underlies our selection of the high-reliability equilibrium as the efficient one.

To better understand Corollary 2 it is helpful to decompose the forces that push a planner away from an equilibrium outcome. Let $x$ be the relationship strength and $r$ the reliability in a hypothetical symmetric equilibrium. Note that the gross profit per functional firm is $g(r)$, so that the gross profit integrated across all firms (functional or nonfunctional) is

$$\mathrm{GP} = rg(r).$$

We can then write net social surplus as follows:

$$\begin{aligned} V &= h(r) - \int_{\mathcal{V}} c(x_v)\,dv \\ &= \left[h(r) - \mathrm{GP}\right] + \left[\mathrm{GP} - \int_{\mathcal{V}} c(x_v)\,dv\right] \\ &= \underbrace{\left[h(r) - rg(r)\right]}_{\mathrm{CS}} + \mathrm{PrS}. \end{aligned}$$

[41] Note that $x^{\mathrm{SP}}(\kappa,\mu)$ is single-valued for such values of $x$, so we may treat $x^{\mathrm{SP}}$ as a function here, abusing notation.

Here we define total producer surplus, PrS, as gross profits net of investment costs. While the term $h(r) - rg(r)$ is, strictly speaking, simply the difference between social welfare and payments made to the firms, it is helpful for intuition to interpret it as consumer surplus. With this decomposition in hand, we can consider a change in the investment level to $x_v + t$. We are interested in the overall wedge between the planner's incentives and firms' incentives. It is helpful to consider how a planner would deviate from the choices that firms make, and hence we evaluate the derivative of $V$ in $t$ at equilibrium. Using the first-order conditions for the firms (see the proof of Lemma SA4 from the online Appendix), the marginal effect on welfare of increasing $t$ from the equilibrium point is

$$\frac{dV}{dt} = \underbrace{\frac{d}{dt}\big[h(r) - rg(r)\big]}_{\text{CS wedge}} + \underbrace{rg'(r)\frac{dr}{dt}}_{\text{business stealing}} + \underbrace{g(r)\frac{dr}{dt}}_{\text{reliability externality}}.$$

There are three terms in the planner's first-order condition that may push the planner away from the equilibrium outcome. First, the planner cares about consumer surplus (the quantity $h(r) - rg(r)$). Firms, on the other hand, do not care about the contribution of reliability to consumer surplus. We term this part of the planner's first-order condition the CS wedge, corresponding to the classic problem of nonappropriability of consumer surplus. Second, there is a business-stealing effect. Producers of a good make lower profits on their sales when there is more competition because reliability is higher (i.e., the unit profit function $g(r)$ is decreasing in $r$). Third, there are reliability externalities: the reliability of a firm's intermediate good production increases the reliability, and hence profitability, of those firms sourcing from it, the firms sourcing from these firms, and so on. Firms do not internalize their contributions to helping *other* varieties function. This pushes a planner to want to choose higher reliabilities, relative to firms.

The novel aspect is the reliability externality. The proof of Corollary 2 analyzes this externality to show that, in any equilibrium, there is always underinvestment. This motivates an interest in policies that increase investment, as this both increases welfare and has the potential to mitigate fragility. We turn to these next.

### B. *Implications for Policies Supporting Investment and Robustness*

Counteracting underinvestment and removing fragility is not straightforward. Here we discuss several intuitive policy interventions that have limited effects, and others that have more potential to qualitatively change the robustness of the system.

As we noted in Section IVC, we view investments as being chosen in the medium run, meaning ex ante of shock realizations. Some policies will affect investment decisions on this timescale. In contrast, a shock causes disruption in the short run, a timescale on which firms cannot adjust their investment decisions. Nevertheless, some policies may be able to react to shocks on this shorter timescale. We will analyze policies in both horizons, beginning with the medium run.

*Marginal Interventions on Investment Incentives.*—Suppose investment in link strength is subsidized to reduce the marginal cost of investments, making the cost of a given investment $1-\theta$ times its original cost. More precisely, we make the gross profit equal to

$$\Pi_{if} = \underbrace{P(x_{if}; x)}_{\text{prob. functional}} \cdot \underbrace{g(r)}_{\substack{\text{gross profit}\\ \text{if functional}}} - \underbrace{\left(\frac{1-\theta}{\kappa}\right)c(y_{if})}_{\text{cost of effort}}.$$

Suppose the status quo is at a critical equilibrium with reliability $r = \underline{r}_{\text{crit}}$ (as in Figure 6 panel B). A subsidy of this form will shift the red best-response curve in Figure 6 panel B to the right. At the margin this will increase reliability, but have a very limited effect on equilibrium investments for high values of the depth parameter $\tau$. Thus the equilibrium will remain fragile. Intuitively, the external subsidy crowds out private strength investment one-for-one in the fragile regime. The same argument can be used for larger values of $\kappa$. This leads to the counterintuitive conclusion that a policy intervention that works well in terms of increasing reliability in normal times can still fail to remove fragility to aggregate shocks.

Indeed, a similar analysis applies for many other interventions that shift the best-response curve to the right—including, for example, improvements in institutions that increase baseline relationship strength $\underline{x}$ marginally.

*Larger Interventions to Improve Investment.*—While marginal interventions typically fail to resolve fragility, nonmarginal interventions might work, for the reasons apparent in the earlier discussion (see also Figure 6 panel B). They can shift the equilibrium into the nonfragile regime from the fragile one. The kinds of policies that would work here are substantial subsidies to investments in strength or actions that nonmarginally increase profitability in an industry. This could be done in a variety of ways: large tax exemptions might be granted or competition might be limited by, for example, improved patent protections.

Such interventions have the flavor of big-push policies advocated for developing counties to escape poverty traps, but are different in a few ways from the standard analysis (Murphy et al. 1989).[42] The standard theory underlying a big-push approach is one of multiple equilibria: there is a low-output equilibrium with traditional production where demand is low and traditional technologies are more cost-effective, and a high-output equilibrium with industrialized production where demand is high and industrial technologies are more cost-effective. For an economy to grow out of the bad equilibrium, it must do so in a balanced way to take advantage of one sector's growth fueling other sectors' demands, beyond external economies of scale that are confined to a given industry (Rosenstein-Rodan 1943). A temporary intervention that subsidizes industrial technology can facilitate this shift. Crucially, the good equilibrium is sustained by increased production under the same institutions. Once a temporary intervention is removed, the technology and other economic primitives remain unchanged: in particular, there is no long-term government support of production, nor are there any institutional changes.

[42] See also Okuno-Fujiwara (1988) and Rodrik (1996).

In our analysis we select undominated equilibria and so there is no scope for policies to facilitate coordination on better equilibria. Had we not made this selection, temporary interventions could shift an economy out of a bad equilibrium and into a good one, just as in big-push theories. Nevertheless, in the fragile regime which is our main focus, all of the productive equilibria involve fragility, and so no temporary policy can remove it. To resolve fragility, permanent intervention on costs or production technologies is necessary.[43] To summarize, in our comparison to classic big-push models we also find an important role for nonmarginal interventions, but for different reasons and with different implications for policy.

*Targeting Interventions*.—Our analysis offers some tentative suggestions for how interventions should be targeted to best address fragility. For targeting questions to make sense, there must be some heterogeneity across the economy. One type of heterogeneity was introduced in the extension on partial knowledge of depth in Section VD. We first consider interventions that will take a supply network out of the fragile regime in the environment presented there. In any productive equilibrium, the reliability of the low-depth firms does not affect the limit reliability of the high-depth firms. Hence, further improvements in reliability for the low-depth firms have no impact on the reliability of high-depth firms. On the other hand, targeting the high-depth firms for nonmarginal interventions such as those discussed above may be able to move the supply network out of the fragile regime.

This relates to the issue of reshoring—bringing production that had been outsourced to foreign suppliers back into a rich country. The rationale for reshoring through the lens of the model is as follows. Links involving long-distance shipping to different countries are more prone to disruption and may occur under weaker institutions. Reshoring can therefore be modeled as improving the strength of relationships among those firms involved or improving the technology for investing in those relationships. The analysis in the previous paragraph suggests that if reshoring targets parts of the supply network that are of low depth (and known to be so by the participating firms), the impact is likely be limited. Reshoring high-depth, complex, products can, on the other hand, be quite effective.

Targeting can also be considered in the model of heterogeneous product networks in the second part of Section VD. There we showed that there is generally a "weakest link" phenomenon in which all firms downstream of a fragile product are also fragile (Proposition 5). Thus, policymakers should be especially interested in identifying upstream, or central, complex industries in a critical equilibrium. We would expect the industries at risk to be the ones with relatively low profit margins and relatively few sourcing options per firm. Beyond these guidelines, a diagnostic indication of being in or near a critical equilibrium is that changes in $\kappa$ (e.g., changes in profitability) are incident only on reliability $\rho$ in equilibrium, as opposed to investment $x$ (as discussed above).

[43] It may be that a more productive economy is capable of supporting institutional reforms or innovations that, for example, change $\kappa$. This is a multiple-equilibria story in a larger model, but is still markedly different from the classic big-push story as described above.

*Reactive Interventions: The Short Run.*—As discussed in Section IVC, the extreme effects of fragility occur in the short run (without giving firms a chance to adjust their investment decisions). Thus, there is a case for quick, reactive short-term government interventions, if such interventions can directly address an aggregate shock. For example, suppose shipping is constrained by regulations that limit the throughput of ports, or short-term shortages of labor in the transport industry. These problems can be modeled as reducing the strength of supply relationships throughout a supply network. Our analysis shows that this can lead to cascading disruptions and production freezes when supply networks are fragile. However, if the government is able to quickly counteract these shocks—e.g., relaxing regulations, or mobilizing military personnel to compensate for the short-term labor shortage—then that can be effective. Moreover, firms' anticipation of such reactions does not significantly change their investment incentives, as long as aggregate shocks are fairly rare and the reactive policy operates only when the shock hits. (The reasoning here is analogous to the discussion of anticipated shocks in Section VC.) Thus, there is no concern of crowding out those incentives. So, perhaps counterintuitively, reactive interventions like this can be more effective in avoiding fragility than the more forward-looking policies discussed above that seek permanent improvements in reliability.

### C. *Embedding Precipices in a Macroeconomic Model: Aggregate Volatility*

So far we have focused on a single complex supply network with particular parameters, interconnected by specific-sourcing relationships. The larger economy can be thought of as consisting of many such supply networks, each one small relative to the economy. Our question in this section is what the fragility of some of these particular supply networks implies for the reaction of the entire economy to shocks. Our main finding is that, while the severe amplification of shocks remains, the conclusion is nuanced by embedding individual supply networks in a richer setting.

Suppose there are many supply networks operating independently of each other, with heterogeneity across supply networks but, for simplicity, homogeneity within each network. The parameters of these different supply networks, including their complexities $m$ and multisourcing numbers $n$, are drawn from a distribution. We know from the above that a small shock to relationship strength can discontinuously reduce the production of some of these supply networks. We now point out that a small shock can have a large macroeconomic effect and that the structure of the fragile regime is essential for this.

For simplicity, fix the function $c(\cdot)$ to be the same throughout the economy.[44] A given supply network is then described by a tuple $\mathfrak{s} = (m, n, \kappa)$. We consider the space of these networks induced by letting the parameters $m, n$, and $\kappa$ vary. In particular, we let $\mathcal{M}$ be the set of possible values of $m$, the set of integers between 1 and $M$; we let $\mathcal{N}$ be the set of possible values of $n$, integers between 1 and $N$, and we allow $\kappa \in \mathcal{K} = [0, K]$. The space of possible supply networks is now

[44] This could also be drawn from a distribution, but the notation would be more cumbersome.

$\mathcal{S} = \mathcal{M} \times \mathcal{N} \times \mathcal{K}$. We let $\Psi$ be a distribution over this space, and assume that it has full support.

In some supply networks there will not exist an equilibrium with positive production, which we henceforth call a productive equilibrium (for example, when $\kappa$ is sufficiently low fixing the other parameters). Consider now those supply networks for which there is a productive equilibrium. There are two possibilities. It may be that the only supply networks for which there is a positive equilibrium have $m = 1$.[45] That is, the only supply networks with positive reliability are simple. In this case, there is no aggregate fragility.

But if, in contrast, $\mathcal{S}$ contains supply networks where production is *not* simple, then we will have macroeconomic fragility. Indeed, an immediate consequence of Theorem 1 is that if there are some complex $(m \geq 2, n \geq 3)$ supply networks with positive equilibria, then some of the lower-$\kappa$ networks with the same $(n, m)$—which are included in $\mathcal{S}$—are in the fragile regime. The measure that $\Psi$ assigns to supply networks in the fragile regime is positive. Thus, a shock to relationship strengths will cause a discontinuous drop in expected aggregate output. On the other hand, in contrast to the situation with a single supply network, aggregate output need not fall all the way down to zero: there will, instead, be a discontinuous loss of some fraction of the output.

So far we have looked at the case in which the different supply networks operate independently and all business-to-business transactions occur through supply relationships confined to their respective supply networks. As we have emphasized, such specific relationships correspond to inputs tailored to the specifications of the business purchasing them. They are not products that can be purchased off-the-shelf. However, many other inputs are sourced in different ways. For example, most business use computers, and buy them off-the-shelf rather than through the specific-sourcing relationships we have focused on. So far we have abstracted from any interdependencies between businesses created by such arm's-length purchases. However, these interdependencies might also be present—indeed, this is the main type of inter-firm dependence found in standard macroeconomic models of production—and they interact in interesting ways with the forces we have studied. If a small aggregate shock causes the collapse of some supply networks, the off-the-shelf inputs available to other supply networks that managed to remain functional become scarcer and more costly. This effectively damages the productivity of these other supply networks, and some of them that were not previously on a precipice will now find themselves there. Thus, they will now be very sensitive to the same aggregate shock that hurt their off-the-shelf suppliers. The key is that being on the precipice is not a fixed attribute of a supply network's structure, but is in fact dependent on its productivity. Thus, even with spillovers that occur through prices of off-the-shelf goods rather than relationships, there can be domino effects where, iteratively, previously robust parts of the economy become fragile.[46]

[45] For example, it might be that $M = 1$ so that only simple production is feasible, or $K$ might be sufficiently low that only simple production has a positive probability of being successful.

[46] In online Appendix SA6 we provide a numerical illustration of this phenomenon.

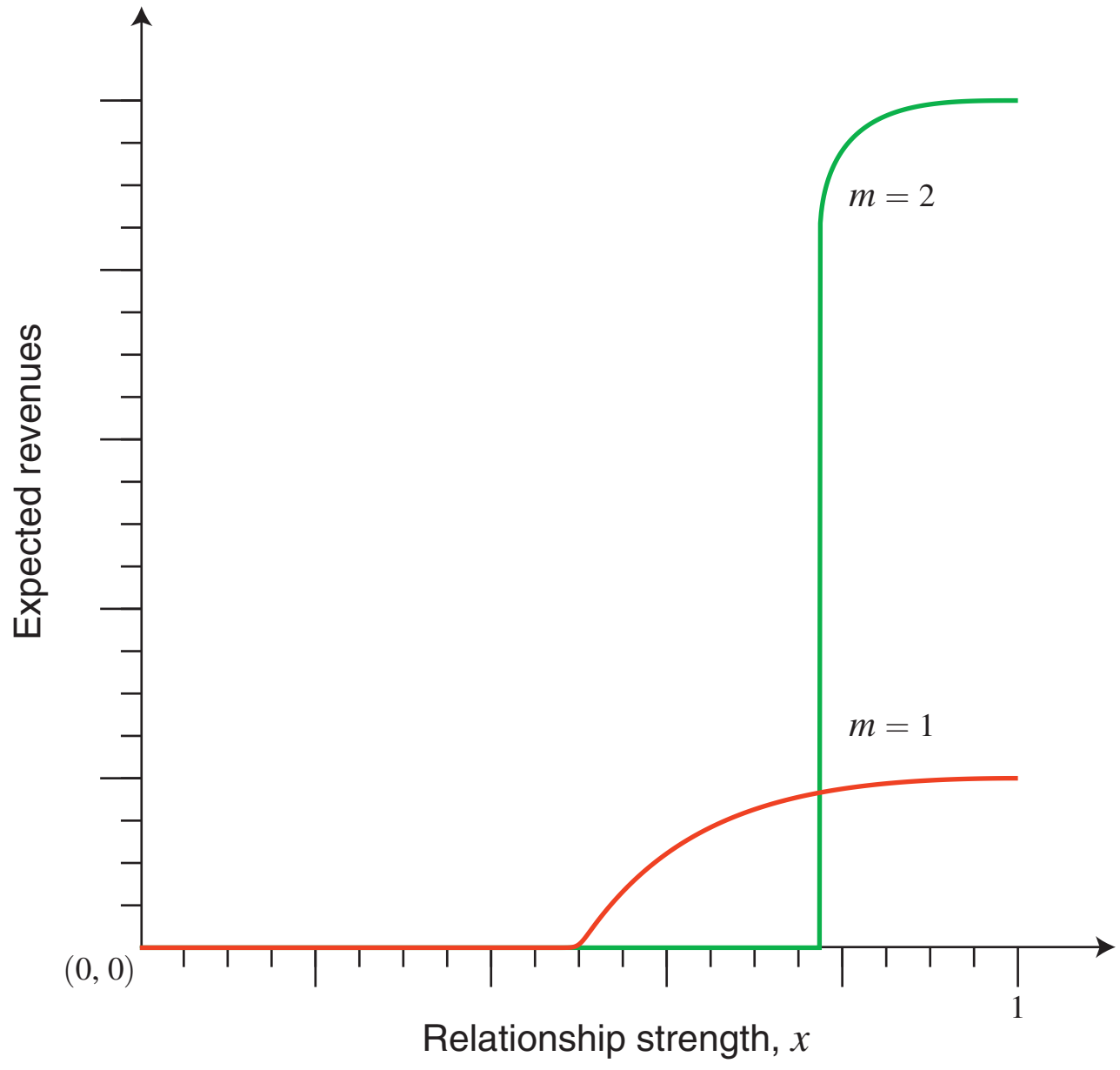

FIGURE 10

*Notes:* A contrast of the $m = 1$ and $m = 2$ cases with the degree of multisourcing being held fixed at $n = 2$. Expected revenues, on the vertical axis, are a product of the probability of successful production and the price of the good. The illustration shows a case in which the complex good retails for a price of 1 while the simple good retails for a price of $1/5$.

## D. *Some Implications for Industrial Development*

The comparison between the production of complex and simple products developed in Section VB has interesting implications for the complexity of technologies used across countries, and for industrial development. While a full analysis of these issues is beyond the scope of this article, Figure 10 illustrates the main point. We posit, similarly to Levine (2012), that the revenue of a product is the associated level of production reliability multiplied by the value of the product, which is greater for more complex products. We then plot the two revenue curves, one derived from complex products (from Figure 3) and one for simple products (from Figure 8 panel B). Small increases in the quality of commercial institutions can make more complex production technologies viable and yield discontinuous benefits to an industry by enabling a transition from simple to complex production.

Even this very rudimentary theory of development and industrialization fits a number of stylized facts: (i) Industrialization, when it occurs, is rapid and economic output increases dramatically. (ii) At the same time, the share of the value of total production that can be attributed to intermediate inputs increases quickly (Chenery, Robinson, and Syrquin 1986). (iii) The quality of institutions, and particularly those related to contracting, can help explain what kinds of production different economies can support (Nunn 2007) and hence cross-country differences in development (Acemoglu and Johnson 2005), wages, and productivities (Jones 2011). (iv) Finally, more complex supply chains are associated with higher rates of disruption (Craighead et al. 2007).[47]

A full study of these issues obviously requires more detailed modeling. Nevertheless the discontinuities we identify may have a useful role to play in theories relating the quality of institutions and an economy's production possibilities.

[47] The unscaled reliability curve for complex production lies below the reliability curve of simple production. So complex production is more valuable but less reliable.

The Supplementary Appendix is available on SSRN.